\documentstyle[preprint,aps,prd]{revtex}

\input{epsf}
\tightenlines

\begin{document}

\newcommand{\eq}{\begin{eqnarray}}
\newcommand{\en}{\end{eqnarray}}
\renewcommand{\theequation}{\arabic{section}.\arabic{equation}}
\newcommand{\mathbold}[1]{\mbox{\boldmath $\bf#1$}}

\preprint{BUTP-01/10}

\draft
\title{Hadronic potentials from effective field theories}

\author{E. Lipartia}
\address{Department of Theoretical Physics 2, Lund University, 
Solvegatan 14A, S22362 Lund, Sweden, and                          
         Laboratory of Information Technologies,
         Joint Institute for Nuclear Research, 141980, Dubna, Russia, and
         HEPI, Tbilisi State University, University St. 9,
         380086 Tbilisi, Georgia}
\author{V. E. Lyubovitskij}
\address{Institute of Theoretical Physics, University of T\"{u}bingen, 
         Auf der Morgenstelle 14, D-72076 T\"{u}bingen, Germany, and
         Department of Physics, Tomsk State University,
         634050 Tomsk, Russia}
\author{A. Rusetsky}
\address{Institute for Theoretical Physics, University of Bern,
         Sidlerstrasse 5, CH-3012, Bern, Switzerland, and
         HEPI, Tbilisi State University, University St. 9,
         380086 Tbilisi, Georgia}

\date{\today}

\maketitle

\begin{abstract}
We construct the potentials that describe the spectrum and decay of
electromagnetic bound states of hadrons, and are consistent with ChPT.
These potentials satisfy the matching condition 
which enables one to express the parameters of the potential through the
threshold scattering amplitudes calculated in ChPT. 
We further analyze the ambiguity in the choice of the short-range 
hadronic potentials, which satisfy this matching condition.

\end{abstract}


\section{Introduction}
\setcounter{equation}{0}
\label{sec:Introduction}

The energy spectrum and decays of electromagnetic bound states of
strongly interacting particles - so called hadronic atoms - have recently 
been measured by several experimental collaborations.  These measurements
yield an extremely valuable piece of information about the behavior of QCD at
a very low energy, which is hardly accessible with a different experimental
technique. In particular, the measurement of the $\pi^+\pi^-$ atom decay 
width by DIRAC collaboration at CERN~\cite{DIRAC} which will result in the
determination of the difference $a_0-a_2$ of the $S$-wave $\pi\pi$ scattering
lengths at a $5\,\%$ precision, would allow one to directly test the 
large/small condensate scenario of chiral symmetry breaking in QCD with two 
flavors. Further, the Pionic Hydrogen collaboration at PSI intends to 
extract the $S$-wave $\pi N$ scattering lengths from the ongoing measurement
of the spectrum and transition energies between $3p-1s$ levels in pionic
hydrogen at a $1\,\%$ accuracy which is unique in hadron physics~\cite{PSI}. This will
yield a more precise value of the $\pi NN$ coupling constant and of the 
$\pi N$ $\sigma$-term. Finally, the DEAR 
collaboration~\cite{DEAR} at the DA$\Phi$NE facility plans to measure the 
energy level shift and lifetime of the $1s$ state in $K^{-}p$ and $K^-d$ 
atoms - with considerably higher precision than in the previous experiment 
carried out at KEK~\cite{KEK} for $K^-p$ atoms. It is expected~\cite{DEAR} 
that this will result in an accurate determination of the $I=0,1$ $S$-wave  
scattering lengths. It will be a challenge for theorists to extract from 
this new information on the $\bar{K}N$ amplitude at threshold a more 
precise value of e.g. the $KN$ $\sigma$-term and of the strangeness content 
of the nucleon.

In order to fully exploit the high-precision experimental data, it is
imperative to design the theoretical framework for the analysis of these
data which would describe this sort of bound states in the accuracy that
matches the experimental precision. In practice, this would mean that the
Deser-type formulae~\cite{Deser} which are used to extract the strong
scattering lengths from the measured values of the energy shift and width of
hadronic atoms, do not provide the necessary precision and should be 
corrected to accommodate strong and electromagnetic isospin-breaking
corrections which typically amount up to a few percent. Since these
corrections arise in a result of a complicated interplay of strong and
electromagnetic effects in the bound-state observables, some dynamical input
on strong interactions is needed for their evaluation. Starting
from~\cite{Trueman}, the problem of the calculation of hadronic atom
observables has been analyzed within the potential scattering theory 
approach (see, e.g.~\cite{Rasche,Rasche-new,Rasche-talk,Sigg}), where the strong 
interactions are described by the short-range ``hadronic'' potentials.
In order to possess a predictive power, certain assumptions about the strong
potential should be made within this approach - it is usually assumed
that the short-range strong potential does not violate the isospin
symmetry and describes purely hadronic data at low
energy. Isospin-breaking corrections are then calculated by 
taking into account the kinematical effect due to the mass difference of
charged and neutral particles, including Coulomb interaction, vacuum
polarization and some higher-order pure QED effects.

The issue of ``purification'' of the experimental data 
with respect to strong and electromagnetic isospin-breaking corrections 
arises in the context of the the low-energy hadronic scattering processes
as well. This problem is closely related to the problem of the hadronic bound
states, which were discussed above. Two distinct approaches are available in
the literature to address the issue in the scattering sector: the dispersion
relations approach~\cite{Tromborg,Pietarinen}, and the potential scattering
theory approach~(see, e.g. \cite{Sauter,Rasche-scat,Zimmermann}).
The results obtained within these two approaches are used e.g. in the 
recent analysis of the low-energy $\pi N$ scattering
data~\cite{Pietarinen,Rasche-analysis}.   

Recently, the problem of isospin-breaking corrections in hadronic atom
observables has been addressed in the framework of ChPT~\cite{Sazdjian-atom,all,Bern1,Bern2,Bern3,Bern4} (see
Ref.~\cite{Bern2} for the comparison of different approaches).
In a result of these investigations, a systematic discrepancy between the
results obtained within the potential scattering theory approach and the 
field-theoretical approach, has been discovered for the case of the
$\pi^+\pi^-$ atom decay width and the $\pi^- p$ atom ground-state energy.
In particular, the prediction for the $\pi^+\pi^-$ atom decay
width~\cite{Rasche,Rasche-new,Rasche-talk} depends on the potential used, and on
the choice of the free Green function in the Schr\"{o}dinger equation.
Earlier calculations~\cite{Rasche} have predicted the isospin-breaking 
corrections with the opposite sign and with the same order of magnitude as in
ChPT. In the later paper~\cite{Rasche-new}, including some kinematic
relativistic 
corrections in the free Green function, the authors obtain a
result for the lifetime $\tau=3.0\times 10^{-15}~sec$
which is still not compatible with the existing prediction of ChPT: 
for a fixed value of $a_0-a_2$, the lifetime
equals to $\tau=2.9\times 10^{-15}~sec$ with an accuracy of
$1.2\,\%$~\cite{Bern4}. The latest prediction of the potential model for the
lifetime is numerically close to that of ChPT~\cite{Rasche-talk}. 
However, no compelling physical reasons are available in
favor of the conclusion, that the latest version includes all relevant 
isospin-breaking corrections from the underlying theory. The
potential-approach based prediction for the energy shift of the $\pi^-p$
atom~\cite{Sigg} differs by a factor $2$ from the order-of magnitude estimate
carried out 
in ChPT at $O(p^2)$~\cite{Bern3}, and quotes the value $0.5\,\%$ for
the systematic error which is $4$ times smaller than the corresponding
value obtained from ChPT. Even if the values of the low-energy
constants (LECs) in ChPT turn out to be such (see Ref.~\cite{Faessler}
for the evaluation of these constants in the quark model),
that the prediction for the
correction in ChPT comes numerically close to that of the potential
approach, still the systematic error is underestimated
in the potential approach. Of course, the large systematic error in the
prediction of ChPT comes from the LECs which, albeit poorly known in the 
$\pi N$ case, are still present and can not be simply disregarded. In
addition, we would like to stress here that there is
no physical reason why the LECs in ChPT must have
the particular values which reproduce the results of the potential approach.
 
The discrepancy mentioned above
does not come to a surprise: the potential which is used to
calculate the bound-state observables, does not include a full content of
isospin-breaking effects in QCD+QED. It has been demonstrated (see,
e.g.~\cite{Zuoz}) that the isospin-symmetric short-range potential 
which has been used so far, can not fully accommodate the effect of the direct
quark-photon interaction, as well as the isospin-breaking effect due to the
explicit dependence of the scattering amplitude on the quark masses which is
governed by the chiral symmetry. These effects turn out to be
dominant in the isospin-breaking part of $\pi\pi$ and $\pi N$ interactions
at threshold, that leads to the above-mentioned discrepancy.
Moreover, since the potential approach makes use of the same ideology
in the calculation of the isospin-breaking corrections to hadronic scattering
amplitudes in the low-energy region, the same criticism is applicable
in this case as well - in fact, a significant discrepancy between the
results of the potential approach and those of the ChPT in
the analysis of $\pi N$ scattering process has already been
reported~\cite{Fettes}. 

The aim of the present paper is to explicitly demonstrate, how the potential
model can be made compatible with ChPT - in what
concerns the calculation of the isospin-breaking corrections to the
observables of hadronic atoms. This construction will help one to finally 
resolve the long-standing puzzle and to determine the status of the potential
approach-based calculations in the analysis of the hadronic atom data. 
Moreover, in the view of the fact that the
potential approach is widely used to evaluate the isospin-breaking effects in
scattering processes, 
it is important to note that the same construction allows one to 
constrain the
potential from comparing the threshold amplitudes in the field theory and in
the potential approach. In this vein, one may hope to reduce the ambiguity in the
prediction of the isospin-breaking corrections above threshold. The ultimate
goal of our investigation is to derive the short-range hadronic 
potentials - including the part which violates the isospin symmetry -
from ChPT. Although
extensive investigations address the similar issue in the context of purely 
strong potential (see, e.g.~\cite{Schrodinger,RG,Kolck}), to the best of our
knowledge, the comprehensive study of isospin-breaking effects is not
available in the literature.

The layout of the present paper is follows. In Section~\ref{sec:strong}
we briefly review the derivation of the potential in the absence of isospin
breaking. Section~\ref{sec:one-channel} deals with the simplified case of
one-channel scattering, whereas in Section~\ref{sec:two-channel} we address
the inclusion of the isospin-breaking effects in full generality
(multichannel scattering, relativistic
corrections). In Section~\ref{sec:Comparison} we compare the results obtained
within ChPT and existing potential models. Finally, 
Section~\ref{sec:Conclusions} contains our conclusions.

\section{Strong potential}
\setcounter{equation}{0}
\label{sec:strong}

For demonstrative purposes,
below we shall briefly review the basic idea which lies in the
foundation of the present approach. Assume that one starts from a 
given relativistic field theory, and is aimed at the construction of the
potential which, used in the Lippmann-Schwinger equation, produces the same
$S$-matrix elements at low energies, as the initial relativistic field
theory. In order to achieve this goal, it has proven very useful, at an 
intermediate stage, to construct the non-relativistic effective field
theory that correctly describes the low-energy behavior of the initial
relativistic theory. To ease the consideration, we
treat here the simplest case of scalar self-interacting field.
Electromagnetic effects are assumed to be absent. In the low-energy
domain (at energies much less than the mass of the particle), the physics
can be described on the basis of the non-relativistic Lagrangian 
\eq\label{L-str}
&&
{\cal L}=\phi^\dagger\biggl(
i\partial_t-M+\frac{\triangle}{2M}+\frac{\triangle^2}{8M^3}+\cdots\biggr)\phi
+g_0(\phi^\dagger\phi)^2+g_1\phi^\dagger\phi^\dagger\stackrel{\leftrightarrow}
{\triangle}\phi\phi+g_2\phi^\dagger\phi\stackrel{\leftrightarrow}
{\triangle}\phi^\dagger\phi+\, ,\cdots
\nonumber\\
&&
\en
where $M$ is the physical mass of the particle, $u\stackrel{\leftrightarrow}
{\triangle}v\doteq u(\triangle v)+(\triangle u) v$, and
$\phi(x)$ is the non-relativistic field operator
\eq\label{phi-nr}
\phi(0,{\bf x})=\int d\nu({\bf p})\,
{\rm e}^{i{\bf p}{\bf x}}\, {\bf a}({\bf p})\, ,\quad\quad
[{\bf a}({\bf p}),{\bf a}^\dagger({\bf k})]
=(2\pi)^3\delta^3({\bf p}-{\bf k})\, ,\quad\quad
d\nu({\bf p})\doteq\frac{d^3{\bf p}}{(2\pi)^3}\, .
\en

The Lagrangian (\ref{L-str}) contains an infinite string of local operators
with an increasing mass dimension, which describe the scattering process
$\phi\phi\rightarrow\phi\phi$. The operators which contain powers of space
derivatives are suppressed by the corresponding power of $M$ which is the 
only heavy scale available. Thus, the contribution of these operators becomes
increasingly suppressed\footnote{We 
consider only the case of the so-called ``natural EFT'' where such an
estimate for the size of the couplings is justified, and one may
straightforwardly apply the perturbation theory to calculate the $S$-matrix
elements.} at a small momenta $|{\bf p}|<<M$.
Note that higher
orders in time derivative $\partial_t$ can be systematically eliminated by 
the use of the equations of motion, without altering the $S$-matrix
elements~(see, e.g.~\cite{Lamb}).

The detailed discussion of the perturbation theory based on the
Lagrangian~(\ref{L-str}) can be found in Refs.~\cite{Bern4,Lamb} -
below, we only provide the sketch of the procedure.
The $S$-matrix element for the scattering process in the non-relativistic
theory is given by a sum of tree diagrams, plus any number of $s$-channel
bubbles, plus mass insertions in the external and internal lines which come
from the higher-order terms in the kinetic part of the Lagrangian (see
Fig.~\ref{fig:strongbubbles}). The free propagator of the scalar field
is given by
\eq\label{propagator}
i\langle 0|T\bar\phi(x)\bar\phi^\dagger(0)|0\rangle=
\int\frac{d^4p}{(2\pi)^4}\,\frac{1}{M+{\bf p}^2/(2M)-p^0-i\epsilon}\, ,
\en
where $\bar\phi$ denotes the free field.
To ease notation, we omit $-i\epsilon$ term in the following.
The elementary building block in the calculation of the diagram with any
number of strong bubbles in the CM frame is given by
\eq\label{bblock}
&&
J(P^0)=\int\frac{d^Dl}{(2\pi)^Di}\,
\frac{1}{M+{\bf l}^2/(2M)-P^0+l^0}\,
\frac{1}{M+{\bf l}^2/(2M)-l^0}\,
=\frac{iM}{4\pi}\,(M(P^0-2M))^{1/2}
\nonumber\\
&&
\en
at $D\rightarrow 4,\,\, P^0>2M$. The scattering matrix elements are
obtained by putting $P^0=2w({\bf p})$ where $w({\bf p})=(M^2+{\bf p}^2)^{1/2}$
is the energy of the particle in the CM frame. At threshold, in
dimensional regularization, the integral is purely imaginary and is
proportional to $i|{\bf p}|$ where ${\bf p}$ is the three-momentum of
the particle in the CM frame. The integrals containing derivative
vertices and/or mass insertions can be considered
analogously. According to the standard power-counting, their
contribution will be suppressed by the corresponding power of $|{\bf p}|$.

The matching condition for the relativistic $T^R$ and non-relativistic
$T^{NR}$ scattering amplitudes in the CM frame reads
\eq\label{match-str}
T^R({\bf p},{\bf q})=4w^2({\bf p})\,T^{NR}({\bf p},{\bf q})\, ,
\quad\quad
|{\bf p}|=|{\bf q}|\, .
\en
This matching condition is understood as follows. At threshold,
both relativistic and non-relativistic amplitudes can be expanded in
powers of CM momenta ${\bf p}$ and ${\bf q}$
\eq\label{expansion}
T^{I}=f_0^{I}+|{\bf p}|f_1^{I}
+{\bf p}^2f_2^{I}+{\bf p}{\bf q}\, f_3^{I}+O({\bf p}^3)\, ,\quad\quad
I=R,NR\, .
\en
The matching condition then gives the relation between the
coefficients of the expansion in the left and right-hand sides of
Eq.~(\ref{match-str}). In the lowest order in ${\bf p}$, the
non-relativistic scattering amplitude is completely determined by the
tree diagram containing the coupling $g_0$
(Fig.~\ref{fig:strongbubbles}a) - bubbles, derivative
vertices and mass insertions all give the contributions that vanish at
threshold. Consequently, the matching condition in the lowest order in
${\bf p}$ reads
\eq\label{match-LO}
4M^2\, (2!)^2\, g_0=T^R({\bf 0},{\bf 0})\, ,
\en
where in the right-hand side the $S$-wave scattering length in the 
relativistic theory appears (here $2!$ is the symmetry factor for the
identical particles in the initial and final states). Note that in the
matching, we have not  
used the perturbative expansion in the relativistic theory, so the
above relation is valid ``in any order in strong coupling constant''. 
Further, using the same technique, the coupling $g_1$ can be related
to the $S$-wave effective radius and the $S$-wave scattering length,
the coupling $g_2$ can be related to the $P$-wave scattering length,
and so on.

In the following discussion, we shall first neglect derivative
couplings, as well as mass insertions ${\bf p}^4/(8M^3)\cdots$.
These will be considered in Section~\ref{sec:two-channel}.
The non-relativistic scattering amplitude in this case is given by
\eq\label{TNR_add}
T^{NR}({\bf p},{\bf q})=\frac{4g_0}{1-\frac{ig_0M}{2\pi}|{\bf p}|}\, .
\en
Now it is straightforward to check that all strong bubbles with
non-derivative vertex  $g_0$ can be resummed with the help of the following
Lippmann-Schwinger equation
\eq\label{LS-str}
T(E;{\bf p},{\bf q})=V({\bf p},{\bf q})
+\frac{1}{2!}\,\int\frac{d^d {\bf l}}{(2\pi)^d}\,V({\bf p},{\bf l})\,
\frac{1}{E-{\bf l}^2/M}\,T(E;{\bf l},{\bf q})\, ,
\en
and
\eq\label{TNR}
({\bf q}|T(E)|{\bf p})\biggr|_{|{\bf q}|=|{\bf p}|,~E={\bf q}^2/M}
=-\, T^{NR}({\bf q},{\bf p})\biggr|_{|{\bf q}|=|{\bf p}|}\, ,
\en
where $V({\bf p},{\bf q})=-(2!)^2\, g_0$, and we have used
dimensional regularization to regulate UV divergences in physical dimensions
$d\rightarrow 3$. Note that the operator $T(E)$ acts in the
Hilbert space of vectors $|{\bf p})$, whereas $T^{NR}({\bf p},{\bf q})$
are the matrix elements of the scattering operator $T^{NR}$ in the Fock space
between two-particle states $|{\bf p},-{\bf p}\rangle={\bf a}^\dagger({\bf p})
{\bf a}^\dagger(-{\bf p})|0\rangle$, with the CM momentum removed (for
details, see~\cite{Bern1,Bern4}). Below, to ease discussion, we shall
not explicitly distinguish between the state vectors in the two spaces.
 
Now we address our main task of deriving the potential from initial
relativistic field theory. In principle, Eq.~(\ref{LS-str}) already solves
this problem - in the dimensional regularization. The potential has then
a $\delta$-type singularity in the coordinate space
\eq\label{V-r}
V({\bf r},{\bf r}')=\int\frac{d^d{\bf p}}{(2\pi)^d}\,
\frac{d^d{\bf q}}{(2\pi)^d}\, {\rm e}^{i{\bf p}{\bf r}-i{\bf q}{\bf r}'}\,
V({\bf p},{\bf q})=-(2!)^2\,g_0\,\delta^d({\bf r}-{\bf r}')\,
\delta^d({\bf r})\, .
\en
In the conventional scattering theory it is customary to deal,
however, not with the pointlike interactions in the dimensional
regularization, but rather with the smooth potentials in the
position space in the physical space dimensions $d=3$. We may
accommodate this feature by smoothing the potential which is
obtained from field theory, and simultaneously adjusting the coupling
$g_0$ so that the scattering amplitude still obeys the matching
condition~(\ref{match-LO}). We demonstrate this procedure for the
simplest case when one assumes the separable ansatz for the resulting
potential
\eq\label{sep-str}
V({\bf r},{\bf r}')\rightarrow V_\beta({\bf r},{\bf r}')=
-(2!)^2\, g_0(\beta)\, v({\bf r})\, v({\bf r}')\, ,\quad\quad
v({\bf r})=\frac{\beta^2{\rm e}^{-\beta r}}{4\pi r}\, ,\quad
v({\bf p})=\frac{\beta^2}{\beta^2+{\bf p}^2}\, ,
\en
where $g_0(\beta)$ stands for the potential strength, and $\beta$ is the
potential range parameter (we anticipate that, in a result of matching,
$g_0(\beta)$ will depend on $\beta$).
The scattering amplitude in the case of the separable interaction is given by
\eq\label{Tsep}
T(E;{\bf p},{\bf q})=-(2!)^2\, g_0(\beta)\, v({\bf p})\, v({\bf q})\, 
\biggl[1+2!\, g_0(\beta)\,
\int\frac{d^3{\bf l}}{(2\pi)^3}\,\frac{v^2({\bf l})}{E-{\bf l}^2/M}
\biggr]^{-1}\, .
\en
The matching condition for the scattering amplitude at threshold then
gives
\eq\label{g0-beta}
4M^2\, (2!)^2\, g_0(\beta)=\frac{T^R({\bf 0},{\bf 0})}
{1+\frac{\beta}{64\pi M}\, T^R({\bf 0},{\bf 0})}\, .
\en
Of course the choice of the separable ansatz is not unique - one may,
e.g. use the local ansatz $V({\bf r},{\bf r}')\rightarrow
\delta^3({\bf r}-{\bf r}')\,V_\beta({\bf r})$ where $\beta$ denotes
a range parameter for the potential. One may further choose any form
for the potential $V_\beta({\bf r})$, e.g. use the double square well
potential~\cite{Rasche}. In the absence of the analytic solution in
this case one will have to solve the matching condition for the
coupling constant numerically. The generalization to higher-order
terms in ${\bf p}^2$ expansion, as well as for higher partial waves is
straightforward. 

To summarize, we note that, according to the point of view adopted
through the present paper, the smooth potentials in the position
space are nothing than a particular regularization of the genuine
pointlike interactions between particles, which stem from the underlying 
field theory. The shape of the potential thus bears no important
physical information. The physical input from the underlying theory is
contained in the couplings, and enters through the matching condition,
which states that the fixed number of terms in the effective-range expansion 
of the scattering matrix elements coincide in the potential approach
and in the underlying relativistic field theory. This is enough for
both approaches to describe the same physics in the vicinity of the
scattering threshold.

\section {Isospin-breaking effects in the one-channel case}
\setcounter{equation}{0}
\label{sec:one-channel}

In this section, we discuss the inclusion of the electromagnetic effects in
the scheme considered above. Once the photons are included, a qualitatively
new feature in the theory emerges: particles with opposite charges can be
bound by Coulomb force at a distances much larger than a typical range of
strong interaction. The observable characteristics of this sort of
bound states - energy levels and decay widths - receive contributions from
strong interactions. This fact enables one to extract strong scattering
lengths from precise measurements of these characteristics, if a
systematic quantitative description of such a bound states is provided 
on the basis of the underlying theory (see,
e.g.~\cite{Bern1,Bern2,Bern3,Bern4}). 

In order to extract the parameters of the strong interactions from
the experimental quantity which contains contributions from both -
electromagnetic and strong - interactions, one has to say explicitly, how 
such a splitting can be systematically performed in the framework of the 
field theory. As a simplest example, one should be able to disentangle
electromagnetic and strong contributions in the mass of a particle which 
takes part in both interactions. In general, the issue has proven to be 
rather subtle - the splitting is convention-dependent. Further we do not 
discuss this question, since it forms a separate subject for the
investigation. Note that in all examples considered below, one may choose the
convention in the underlying relativistic theory so that the mass of the
particle in the ``purely strong'' world coincides with the mass of charged
particle in the real world, and the explicit prescription for the splitting
of the scattering amplitude into the strong part and the 
electromagnetic correction can be provided. Below, we shall follow these
conventions.

The question which we investigate here can be now formulated as follows.
We study the observables of electromagnetic bound states - energy and
decay width in a theory where both strong and electromagnetic interactions
are present. In the leading order in fine structure constant $\alpha$, the
relation between these observables and the strong scattering lengths is
universal~\cite{Deser}. Nontrivial interplay between electromagnetic and
strong effects occurs at the first non-leading order in $\alpha$. Our aim
is to evaluate these observables at the first non-leading order, and to
construct the short-range hadronic potential which, when used in the
Schr\"{o}dinger equation, leads to the same values of these observables at 
the same order in $\alpha$.

\subsection{Energy-level shift in field theory}

The relativistic field-theoretical Lagrangian of the model which is 
considered in the present Section, describes 
the charged scalar particle with the mass $M$, interacting with photons.
In addition, the Lagrangian includes arbitrary self-interactions of the 
scalar particle.
In the theory, there are two-particle loose bound states with zero total 
charge, which are formed mainly by the static Coulomb interaction.
In the lowest order in $\alpha$, the energy of the ground state coincides 
with the ground-state energy in the pure Coulomb potential:
$E_0=2M-\frac{1}{4}\,\alpha^2M$, and, as was mentioned above,
 there are higher-order corrections to this
value, caused both by the electromagnetic and strong interactions. 
Our goal is to calculate the energy of
the ground state up to and including $O(\alpha^4)$, 
where both these sources contribute.
It is both convenient and conventional to split the ground-state energy at
$O(\alpha^4)$ in the electromagnetic part and strong shift, according
to~\cite{Sigg,Bern3} 
\eq\label{splitting}
E^{\rm tot}=E^{\rm em}+\Delta E^{\rm strong}
\en
The expression for the electromagnetic part can be easily obtained,
adapting formulae from Ref.~\cite{Bern3} to the scalar case\footnote{It
  suffices to substitute $\lambda=\frac{1}{3}\,\langle r^2\rangle$ in Eq.~(4)
  of Ref.~\cite{Bern3} and evaluate energy shift according to the formulae
  (13) and (14) at $m_p=M_{\pi^+}=M$. 
  Vacuum polarization contribution is omitted.}
\eq\label{E-em} 
E^{\rm em}=E_0-\frac{13}{64}\, M\alpha^4
+\frac{1}{6}\, M^3\langle r^2\rangle\alpha^4+o(\alpha^4)\, ,
\en
where $\langle r^2\rangle$ denotes the charge radius of the particle.
Assuming $\langle r^2\rangle=0$, we reproduce the result for the ground-state
energy-level shift at $O(\alpha^4)$, given in Ref.~\cite{Nandy}.

In the following, we shall concentrate exclusively on the strong shift at 
$O(\alpha^4)$ which is given by the second term in Eq.~(\ref{splitting}).

\subsubsection{Non-relativistic Lagrangian}

The non-relativistic Lagrangian which is sufficient to calculate the strong
shift at $O(\alpha^4)$, 
consists of the non-relativistic kinetic term, static Coulomb
interaction, and the short-range interaction described by the local
four-particle non-derivative vertex~\cite{Bern3}
\eq\label{Lagrangian}
{\cal L}&=&{\cal L}_0+{\cal L}_C+{\cal L}_S\, ,\nonumber\\[2mm]
{\cal L}_0&=& \sum_{i=\pm}\phi_i^{\dagger}\left (i\partial_t-M +
\frac{\triangle}{2 M}\right )\phi_i+\cdots\, ,\nonumber \\[2mm]
{\cal L}_C&=& -4\pi\alpha\, (\phi_-^{\dagger}\phi_-)\triangle^{-1}
(\phi_+^{\dagger}\phi_+)\, ,\nonumber\\[2mm]
{\cal L}_S&=&c\, (\phi_+^{\dagger}\phi_-^{\dagger})(\phi_+\phi_-)+\cdots\, ,
\en
where the non-relativistic fields $\phi_\pm$ describe the pair of charged
particles. All higher-order derivative terms that one may write here, do not
contribute to the strong shift at the accuracy we are working (we refer the
reader to Ref.~\cite{Bern3} to more details). 
Further, the coupling constant $c$ in Eq.~(\ref{Lagrangian}) is
determined from matching the relativistic and non-relativistic scattering 
amplitudes at threshold (see below). Generally, one may write
$c=c_0+\alpha c_1+O(\alpha^2)$, where $c_0$ is obtained by matching
to the strong relativistic Lagrangian in the absence of
electromagnetic interactions.

On the potential scattering theory language, the non-relativistic
Lagrangian~(\ref{Lagrangian}) describes one-channel scattering problem for
Coulomb + strong interactions, in the channel with total charge $Q=0$.
The only difference is, that now the strong coupling $c$ - through 
matching to the underlying relativistic theory - depends on $\alpha$ as well.
This is the way how the full content of isospin-breaking effects in the
initial theory enter in the non-relativistic framework.

\subsubsection{Hamiltonian framework and the energy of ground state}

It is convenient to use the Hamiltonian
formulation of the non-relativistic theory. 
The Hamiltonian, derived from the Lagrangian~(\ref{Lagrangian}) 
with the use of canonical formalism, is given by
\eq\label{Hamiltonian}
{\bf H}&=&{\bf H}_0+{\bf H}_C+{\bf H}_S\, ,\quad\quad
{\bf H_{\Gamma}}=\int\,d^3{\bf x}\,{\cal H}_{\Gamma}(0,{\bf x})\, ,\quad
\Gamma=0,C,S
\nonumber\\[2mm] 
{\cal H}_0 &=&\sum_{i=\pm}\phi_i^{\dagger}\left (M-
\frac{\triangle}{2M}\right )\phi_i\nonumber\\ [2mm]
{\cal H}_C &=&4\pi\alpha\, (\phi_-^{\dagger}\phi_-)
{\triangle}^{-1}(\phi_+^{\dagger}\phi_+)\nonumber\\[2mm]
{\cal H}_S &=& - c\, (\phi_+^{\dagger}\phi_-^{\dagger})(\phi_+\phi_-)
\en

The scattering states in the non-relativistic theory are
$|{\bf P},{\bf p}\rangle=
a_+^{\dagger}({\bf p}_1) a_-^{\dagger}({\bf p}_2)|0\rangle$, 
where ${\bf P}={\bf p}_1+{\bf p}_2$
and ${\bf p}=\frac{1}{2}\,({\bf p}_1-{\bf p}_2)$ are the CM and relative 
momenta, respectively. The CM momentum is removed from the
matrix elements of any operator $R$ in Fock space by using the notation
\eq\label{CM}
\langle{\bf P, q}|R(z)|{\bf 0, p}\rangle=
(2\pi)^3\delta^3({\bf P})({\bf q}|r(z)|{\bf p}),
\en
The operator $r(z)$ acts in the Hilbert space of vectors 
$|{\bf p})$, where the scalar product is defined as the integral over the 
relative three-momenta of particle pairs
\eq
({\bf q}|{\bf p})=(2\pi)^3\,\delta^3({\bf q}-{\bf p})\, ,\quad\quad
{\bf 1}=\int d\nu({\bf p})|{\bf p})({\bf p}|\, .
\en 

The Hamiltonian~(\ref{Hamiltonian}) at $c=0$ has an infinite number of
pure Coulomb discrete eigenstates. The ground-state eigenvector in an arbitrary
reference frame is given by
\eq
|\Psi_0,{\bf P}\rangle=\int\frac{d^3{\bf q}}{(2\pi)^3}\Psi_0({\bf q})
|{\bf P},{\bf q}\rangle\, ,
\en
where $\Psi_0({\bf p})$ is the Coulomb wave function in the momentum space
\eq
({\bf p}|\Psi_0)\doteq\Psi_0({\bf p})
=\frac{(64\pi\gamma^5)^{1/2}}{({\bf p}^2+\gamma^2)^2}\, ,\quad\quad
\gamma=\frac{1}{2}\,\alpha M\, .
\en

The eigenstate $|\Psi_0,{\bf P}\rangle$ satisfies the following bound-state
equation 
\eq
({\bf H}_0+{\bf H}_C)\, |\Psi_0,{\bf P}\rangle
=\biggl( E_0+\frac{{\bf P}^2}{4 M^2}
\biggr)\,|\Psi_0,{\bf P}\rangle\, , \quad\quad
E_0=2M-\frac{1}{4}\, M\alpha^2\, .
\en

The resolvent in pure Coulomb theory is defined by 
${\bf G}_C(z)=(z-{\bf H}_0-{\bf H}_C)^{-1}$. In the CM frame this resolvent
is given (cf with Eq.~(\ref{CM})) 
by the Schwinger's representation~\cite{Schwinger}
\eq\label{schw}
({\bf q}| {\bf g}_C(z)|{\bf p})&=& 
\frac{(2\pi)^3\delta^{3}({\bf q}-{\bf p})}
  {E-{\bf q}^2/M}-\frac{1}{E-{\bf q}^2/M}\,
  \frac{4\pi\alpha}{|{\bf q}-{\bf p}|^2}\,\frac{1}{E-{\bf p}^2/M}
  \nonumber\\[2mm]
  &-& \frac{1}{E-{\bf q}^2/M}\,\,
  4\pi\alpha\eta I(E;{\bf q},{\bf p})\,\,\frac{1}{E-{\bf p}^2/M}\, ,
\en
with
\eq\label{Schwinger-I}
I(E;{\bf q},{\bf p})=
\int_0^1\frac{x^{-\eta} dx}
  {[({\bf q}-{\bf p})^2x+\eta^2/\alpha^2(1-x)^2(E-{\bf q}^2/M)
    (E-{\bf p}^2/M)]}\, ,
\en
where $\eta=\frac{1}{2}\,\alpha\,(-E/M)^{-1/2}$ and
$E=z-2M$.

When $c\neq 0$, the bound-state poles in the full resolvent
${\bf G}(z)=(z-{\bf H}_0-{\bf H}_C-{\bf H}_S)^{-1}$ are shifted from their
original Coulomb values. Applying Feshbach's formalism~\cite{Feshbach},
we obtain the equation for the shifted ground-state pole
position~\cite{Bern1,Bern3,Bern4,Lamb}.  
\eq\label{Pole_position}
z-E_0-(\Psi_0|\tau(z)|\Psi_0)=0
\en
where $\tau(z)$ obeys the Lippmann-Schwinger equation
\eq\label{tau-f}
\tau(z)={\bf h}_S+{\bf h}_S \bar{\bf g}_C(z)\tau(z)\, ,\quad\quad
({\bf q}|\bar{\bf g}_C(z)|{\bf p})=({\bf q}|{\bf g}_C(z)|{\bf p})
-\frac{({\bf q}|\Psi_0)(\Psi_0|{\bf p})}{z-E_0}\, .
\en
Here $\bar{\bf g}_C(z)$ denotes the pole-subtracted Coulomb Green function,
and $({\bf q}|{\bf h}_S|{\bf p})=-c$.

The bound-state equation~(\ref{Pole_position}), together with
Eq.~(\ref{tau-f}),  can be solved iteratively,
along the similar lines as in Refs.~\cite{Bern1,Bern3,Bern4}.
We use the dimensional regularization both for ultraviolet and infrared
divergences. The key observation is, that in dimensional regularization,
the subsequent terms in the iterative solution of Eq.~(\ref{tau-f}) for the
quantity $\tau(z)$ are suppressed by powers of $\alpha$ (roughly, one may
count ${\bf h}_S\sim 1,\,\bar{\bf g}_C\sim\alpha$), so that only
a finite number of iterations survives in a given order in $\alpha$. At 
$O(\alpha^4)$, we obtain
\eq
z-E_0=-\frac{\alpha^3M^3}{8\pi}\, c\left\{1-\frac{\alpha M^2}{8\pi}\, c \left (\Lambda(\mu)+
\ln\frac{M^2}{\mu^2}+2\ln\alpha-3\right )\right\}+o(\alpha^4)\, ,
\en
where
\eq
\Lambda(\mu)=(\mu^2)^{d-3}\left(\frac{1}{d-3}-\Gamma'(1)-\ln4\pi\right)\, ,
\en
and $d\rightarrow 3$ in physical space dimensions.

We define the renormalized coupling constant $c^r(\mu)$ as
\eq
c=c_0+\alpha c_1+O(\alpha^2)=
c^r(\mu)+\frac{\alpha M^2}{8\pi}\,[c^r(\mu)]^2\Lambda(\mu)+O(\alpha^2)\, .
\en
Note that both $c_0$ and $c_1$ can be assumed to be real. 
At the lowest order in fine structure constant, the imaginary part of $c$
is determined by the process $\phi_+\phi_-\rightarrow 2\gamma$ in the 
relativistic theory and starts at $O(\alpha^2)$.

Expressed in terms of $c^r(\mu)$,
the energy-level shift does not contain the ultraviolet divergence
\eq\label{DeltaE}
&&\Delta E={\rm Re}\,z-E_0=
-\frac{\alpha^3M^3}{8\pi}\, {\rm Re}\, c^r(\mu)\left\{1
-\frac{\alpha M^2}{8\pi}\,{\rm Re}\, c^r(\mu)\,\left (
\ln\frac{M^2}{\mu^2} +2\ln\alpha-3 \right )\right\}+o(\alpha^4)\, .
\nonumber\\
&&
\en
It is easy now to check that $\Delta E$ does not depend on $\mu$ at
$O(\alpha^4)$.

\subsubsection{Matching to the relativistic amplitude}

Above, we have expressed the strong energy-level shift of the ground state in
terms of the renormalized coupling constant $c^r(\mu)$ in the non-relativistic
Lagrangian. In order to have connection to the initial relativistic
theory, one has to express this coupling - through the matching of
relativistic and non-relativistic scattering amplitudes at threshold - via the
parameters of the relativistic Lagrangian. From Eq.~(\ref{DeltaE}), 
one may
conclude that $c^r(\mu)$ should be known at $O(\alpha)$, so that it suffices
to match the amplitudes calculated at the same accuracy. Further, note that, 
in order to obtain the
strong shift, one has to perform matching for truncated amplitudes
obtained by discarding
(both in the relativistic and non-relativistic theories)
all diagrams that are made disconnected by cutting one photon line:
$\tilde T=T-T_{ex}$, where $T_{ex}$ stands for the one-photon exchange
contribution~\cite{Bern3}. The contribution from $T_{ex}$ is contained
in the electromagnetic shift~(\ref{E-em}).

The matching condition for truncated amplitudes in the CM frame 
is analogous to Eq.~(\ref{match-str}) in the strong sector
\eq\label{match-1c}
\tilde T^R({\bf q},{\bf p})=4w^2({\bf p})\, 
\tilde T^{NR}({\bf q},{\bf p})\, .
\en

In the non-relativistic theory, the full scattering amplitude
at $O(\alpha)$ in the vicinity of threshold is determined by the
diagrams depicted in Fig.~\ref{fig:NR-1c} which are calculated by using the
Lagrangian~(\ref{Lagrangian}). The truncated amplitude $\tilde T^{NR}$ is
obtained by discarding the contribution coming from diagram in
Fig.~\ref{fig:NR-1c}a which corresponds to the exchange of the Coulomb photon.
Further, the non-relativistic amplitude contains the (divergent) Coulomb phase
which is removed. Finally, the expression for the real part of the
non-relativistic truncated amplitude at threshold is given by
\eq\label{NR-1c-thr}
{\rm Re}\,\bigl[{\rm e}^{-2i\alpha\theta_C}
\tilde T^{NR}({\bf q},{\bf p})\bigr]\,
\biggr|_{|{\bf q}|=|{\bf p}|\rightarrow 0}=
\frac{B_1^{NR}}{|{\bf p}|}+B_2^{NR}
\ln\frac{2|{\bf p}|}{M}+{\cal A}^{NR}+O(|{\bf p}|)\, ,
\en
where
\eq\label{amp-NR}
&&B_1^{NR}=\frac{\alpha\pi M}{2}\, {\rm Re}\, c^r(\mu)+o(\alpha)\, ,\quad\quad
B_2^{NR}=-\frac{\alpha M^2}{4\pi}\, [{\rm Re}\, c^r(\mu)]^2+o(\alpha)\, ,\quad\quad
\nonumber\\[2mm]
&&{\cal A}^{NR}={\rm Re}\, c^r(\mu)\,\biggl\{1
-\frac{\alpha M^2}{8\pi}\, {\rm Re}\, c^r(\mu)\,
\biggl(\ln\frac{M^2}{\mu^2}-1\biggr)\biggr\}+o(\alpha)\, ,
\en
and
\eq\label{theta}
\theta_C=\frac{M}{2 |{\bf p}|}\,\left\{\mu^{d-3}\left (\frac{1}{d-3}-
\frac{1}{2}\, (\Gamma'(1)+\ln 4\pi)\right )+\ln\frac{2|{\bf p}|}{\mu}\right\}
\en
The scattering amplitude in the relativistic theory contains the diagrams that
can be made disconnected by cutting one virtual photon
line~Fig.~\ref{fig:R-1c}a, and the rest which, by definition, coincides with
truncated amplitude Fig.~\ref{fig:R-1c}b. After the removal of Coulomb
phase, the truncated relativistic amplitude, like its non-relativistic
counterpart, in the vicinity of threshold contains singular terms
\eq\label{R-1c-thr}
{\rm Re}\,\bigl[{\rm e}^{-2i\alpha\theta_C}
\tilde T^{R}({\bf q},{\bf p})\bigr]\,
\biggr|_{|{\bf q}|=|{\bf p}|\rightarrow 0}=
\frac{B_1^{R}}{|{\bf p}|}+B_2^{R}
\ln\frac{2|{\bf p}|}{M}+{\cal A}^{R}+O(|{\bf p}|)\, .
\en
The matching condition~(\ref{match-1c}) for the regular parts of the
amplitudes at threshold gives
\eq\label{regular-1c}
{\cal A}^R=4M^2\, {\cal A}^{NR}\, .
\en
Using Eq.~(\ref{amp-NR}), one may relate $c^r(\mu)$ with ${\cal A}^R$
\eq\label{cr-A}
{\rm Re}\, c^r(\mu)=\frac{1}{4M^2}\,{\cal A}^R\,\biggl\{
1+\frac{\alpha}{32\pi}\,{\cal A}^R\,
\biggl(\ln\frac{M^2}{\mu^2}-1\biggr)\biggr\}+o(\alpha)
\, .
\en
Finally, substituting this relation into the expression for the strong
energy-level shift, we get
\eq\label{energyshift}
\Delta E=-\frac{\alpha^3M}{32\pi}\,{\cal A}^R\,\biggl\{
1-\frac{\alpha}{16\pi}\,{\cal A}^R\,(\ln\alpha-1)\biggr\}+o(\alpha^4)\, ,
\en
where reference to the non-relativistic theory has completely disappeared -
the strong shift is expressed in terms of the regular part of the relativistic
scattering amplitude at threshold.

\subsubsection{Matching condition for the non-relativistic couplings}

As pointed out above, we implicitly assume the prescription which enables one
to split the scattering amplitude into the strong piece and the
electromagnetic correction which is proportional to $\alpha$
\eq\label{exp-1c}
{\cal A}^R={\cal A}^R_0+\alpha\,{\cal A}^R_1+O(\alpha^2)\, .
\en
Then, from the matching condition~(\ref{cr-A}), for the non-relativistic
coupling we get
\eq\label{c}
c_0=\frac{1}{4M^2}\, {\cal A}^R_0\, ,\quad
c_1=\frac{1}{4M^2}\,\biggl\{{\cal A}^R_1+\frac{1}{32\pi}\,
\bigl[{\cal A}^R_0\bigr]^2\,\biggl(\Lambda(\mu)+\ln\frac{M^2}{\mu^2}-1\biggr)
\biggr\}\, ,
\en
where $c=c_0+\alpha c_1+O(\alpha^2)$.
Note that the equations for determining the energy of the bound state are
formally the same in the non-relativistic effective Lagrangian approach and 
in the non-relativistic potential model. This again, as in the purely
strong case, allows one to interpret
the local 4-particle interaction in the Hamiltonian~(\ref{Hamiltonian})
as the contact potential $V({\bf r})=-c\,\delta^d({\bf r})$ with the strength
determined from matching to the relativistic field theory~(\ref{c}).
If one uses the dimensional regularization throughout, this is a perfectly
consistent interpretation - we have seen that the ultraviolet divergence
contained in $c$ cancels with the divergence arising in the bound-state
calculations. We see also that the coupling $c$ contains the piece
proportional to $\alpha$ which is related to the corresponding piece
$\alpha\,{\cal A}^R_1$ in the expression of the relativistic amplitude.
This means that, in general, one can not assume that the contact term
corresponds to purely strong interactions.
 
Below, we shall demonstrate that, in a complete analogy with a purely strong
case,  one may also construct
conventional smooth short-range potential that reproduces the answer for the
energy of the ground state, obtained within the field-theoretical approach.

\subsection{Energy-level shift in potential model: universality}

Below, we consider the non-relativistic potential model with the potential
given by a sum of Coulomb and short-range parts
\eq\label{v}
({\bf q}|v|{\bf p})\doteq v({\bf q},{\bf p})=
-\frac{4\pi\alpha}{|{\bf q}-{\bf p}|^2}+u({\bf q},{\bf p})=
({\bf q}|{\bf h}_C|{\bf p})+u({\bf q},{\bf p})\, .
\en

One possible interpretation of this potential is to be the
ultraviolet-regulated version of the contact potential introduced in the
previous section, to which it reduces in the local limit
$u({\bf q},{\bf p})\rightarrow -\, c$. According to the discussion above,
the short range potential, in general, should contain the strong 
part, as well as the piece proportional to $\alpha$: 
$u({\bf q},{\bf p})=u_0({\bf q},{\bf p})+\alpha u_1({\bf q},{\bf p})
+O(\alpha^2)$. Our goal is to construct the short-range potential 
$u({\bf q},{\bf p})$ which reproduces the result~(\ref{energyshift}) for the
energy shift of the ground state, obtained within field theory. 
One would expect, that to this end it is necessary to perform matching of
scattering amplitudes at threshold according to the same matching
condition~(\ref{match-1c}) as in field theory. This should result into the
matching condition between the relativistic threshold amplitude ${\cal A}^R$
and the short-range potential $u({\bf q},{\bf p})$ which resembles
the matching condition~(\ref{c}) for the non-relativistic couplings.

Below, we shall demonstrate that the above conjecture indeed holds for the
model considered here. To be more specific, one has to prove the 
universality of the relation~(\ref{energyshift}):
{\em The relation between the energy shift of the ground state and the 
threshold scattering amplitude at next-to-leading order in $\alpha$ is the 
same in field theory and in the potential model.}

In order to prove this statement, we look for the ground state pole in the
scattering matrix which obeys the Lippmann-Schwinger equation
\eq\label{LS}
T(z)=v+v\, {\bf g}_0(z)T(z)\, ,\quad\quad
({\bf q}|{\bf g}_0(z)|{\bf p})=\frac{(2\pi)^3\delta^3({\bf q}-{\bf p})}
{E-{\bf q}^2/M}\, .
\en

The relation of $T(z)$ to the scattering amplitude $T^{NR}({\bf q},{\bf p})$ 
is given by Eq.~(\ref{TNR}).
The shift in the ground state pole position in the potential model is again
given by Eq.~(\ref{Pole_position}), where $\tau(z)$ now satisfies the
equation~(\ref{tau-f}) with $({\bf q}|{\bf h}_S|{\bf p})=u({\bf q},{\bf p})$.

An important remark is in order. In the non-relativistic effective field
theory, we have used dimensional regularization to handle ultraviolet
divergences. In this scheme, one may solve the equation for $\tau(z)$
iteratively, since, as it is easy to see, higher-order iterations amount to
higher-order contributions in $\alpha$ to the bound-state energy.
This is no longer the case for the short-range potential 
$u({\bf q},{\bf p})$: iterations corresponding to the free propagation of
particles should be summed up in all orders. In order to do this, note that
the Schwinger's representation for Coulomb Green~(\ref{schw}) function 
suggests its decomposition into 0-Coulomb, 1-Coulomb and many-Coulomb pieces.
Consequently,
\eq\label{decomp}
\bar {\bf g}_C(z)={\bf g}^{0-C}(z)+{\bf g}^{1-C}(z)+\bar {\bf g}^{n-C}(z)
={\bf g}_0(z)+\delta {\bf g}(z)\, .
\en 
Whereas in the above decomposition $\delta {\bf g}(z)$ which contains the
exchanges of at least one photon can be still treated as a perturbation, one
has to sum all iterations of ${\bf g}_0(z)$. To this end, we introduce the
auxiliary scattering matrix
\eq\label{def-t}
t(z)=u+u\,{\bf g}_0(z)\, t(z)\, .
\en

The relation between $\tau(z)$ and $t(z)$ is given by
\eq\label{tau-t}
\tau(z)=t(z)+t(z)\,\delta {\bf g}(z)\,\tau(z)\, .
\en

The partial-wave expansion of the potential and scattering
matrix is given by
\eq\label{partial}
({\bf q}|u|{\bf p})&=&4\pi\sum_{LM}Y_{LM}(\hat{\bf q})\, u_L(q,p)\,
Y_{LM}^{\star}(\hat{\bf p})\, ,
\nonumber\\[2mm]
({\bf q}|t(z)|{\bf p})&=&4\pi\sum_{LM}Y_{LM}(\hat{\bf q})\, t_L(z;q,p)\,
Y_{LM}^{\star}(\hat{\bf p})\, ,
\nonumber\\[2mm]
({\bf q}|\tau(z)|{\bf p})&=&4\pi\sum_{LM}Y_{LM}(\hat{\bf q})\, \tau_L(z;q,p)\,
Y_{LM}^{\star}(\hat{\bf p})\, ,
\en
where $q=|{\bf q}|$, $p=|{\bf p}|$ and $\hat{\bf q}$ and $\hat{\bf p}$ denote
unit vectors in the direction of  ${\bf q}$ and ${\bf p}$, respectively, and
$Y_{LM}(\hat{\bf q}) \doteq (4\pi)^{-1/2} \,\langle \hat{\bf q}|LM\rangle$. 
Since the ground state contains only $S$-wave, only the term with $L=0$
counts in these sums. Hereafter, we shall suppress the index $L$ in the
partial wave amplitudes. 

Below we assume, that potential $u({\bf q},{\bf p})$ is short-ranged,
hermitian and real. In this case, the unitarity condition for the scattering
matrix $t(z)$ in the $S$-wave then takes the form
\eq\label{unitarity}
{\rm Im}\, t(z;q,p)=-\frac{M k_0}{4\pi}\, t(z;q,k_0)\,
t^\star(z;k_0,p)\, ,\quad\quad k_0=\sqrt{M E}\, .
\en

Now, solving Eq.~(\ref{tau-t}) by iterations, for the energy shift we obtain
\eq\label{dE}
\Delta E&=&{\rm Re}\,(\Psi_0|t(E_0)|\Psi_0) 
+{\rm Re}\,(\Psi_0|t(E_0)\,{\bf g}^{1-C}(E_0)\, t(E_0)|\Psi_0)
\nonumber\\[2mm]
&+&{\rm Re}\,(\Psi_0|t(E_0)\,\bar{\bf g}^{n-C}(E_0)\,
t(E_0)|\Psi_0)+o(\alpha^4)\, .
\en

Below, we proceed with the evaluation
of all three terms in the right-hand side of Eq.~(\ref{dE}):

{\bf 1.} The term with no Coulomb exchanges is given by
\eq
&&{\rm Re}\,(\Psi_0|t(E_0)|\Psi_0)=\phi_0^2\,{\rm Re}\,\,
t(E_0;0,0)\nonumber\\[2mm]
&&+2\phi_0\int d\nu({\bf q})
\Psi_0({\bf q})\, \left ({\rm Re}\,\,t(E_0;q,0)
-{\rm Re}\,\,t(E_0;0,0)\right )+
o(\alpha^4)\, ,
\en
where $\phi_0=\int d\nu({\bf q})\Psi_0({\bf q})=(\gamma^3/\pi)^{1/2}$. 
Using the following 
property of the scattering matrix, that can be proven for short-range
potentials~(see Appendix~\ref{app:phi})
\eq\label{phi0}
\lim\limits_{q\to 0}\frac{1}{q}\,\left ( {\rm Re}\, t(E_0;q,0)-
{\rm Re}\, t(E_0;0,0)\right )=0\, ,
\en
this term can be rewritten as
\eq
{\rm Re}\, (\Psi_0|t(E_0)|\Psi_0)=
\phi_0^2\left\{{\rm Re}\, t(E_0;0,0)+
\frac{4 M\alpha}{\pi}\,{\cal Q}[t]\right\}+o(\alpha^4)\, .
\en
Here $\bar{t}\doteq{\rm Re}\, t(2M;0,0)$ and
\eq
{\cal Q}[t]=\int\limits_0^{\infty}\frac{d q}{q^2}\left 
({\rm Re}\, t(2M;q,0)-\bar{t}\right )\, .
\en
Note that we have substituted $E_0=2M$ in the correction term, since this 
does not affect the result at the accuracy we are working.

Next, we have to perform an analytic continuation of the scattering matrix
from $E_0$ to $2M$. The distance between these two points is of 
order $\alpha^2$. However, since the real part of the scattering matrix
has the unitary cusp at threshold, the difference between 
$t(E_0)$ and $t(2M)$ is of order $\alpha$ rather than $\alpha^2$.
It is easy to check that
\eq\label{cusp}
{\rm Re}\,t(E_0;0,0)-\bar{t}=\frac{\alpha M^2}{8\pi}\, \bar{t}^{\,\, 2}+
o(\alpha)
\en

Collecting all contributions, we finally get
\eq\label{0-C}
{\rm Re}(\Psi_0|t(E_0)|\Psi_0)=\phi_0^2\left\{\bar{t}+
\frac{\alpha M^2}{8\pi}\bar{t}^{\,\, 2}+
\frac{4\alpha M}{\pi}{\cal Q}[t]\right \}+o(\alpha^4)\, .
\en

{\bf 2.} In the calculation of the matrix element corresponding to 1 
Coulomb photon exchange, it is convenient to separate the contributions 
coming from small and large integration momenta. This can be achieved
by rewriting the Coulomb potential as
\eq\label{separ}
\frac{1}{|{\bf q}-{\bf p}|^2}=\frac{1}{|{\bf q}-{\bf p}|^2+b^2}
+\biggl\{\frac{1}{|{\bf q}-{\bf p}|^2}
-\frac{1}{|{\bf q}-{\bf p}|^2+b^2}\biggr\}\, .
\en

The integration can be straightforwardly  carried out, resulting in
\eq\label{1-C}
(\Psi_0|t(E_0){\bf g}^{1-C}(E_0)
t(E_0)|\Psi_0)=
\phi_0^2\left\{-\frac{\alpha M^2}{4\pi}\bar{t}^{\,\, 2}\ln\frac{b}{2\gamma}+
\frac{\alpha M^2}{4\pi^3}{\cal R}[b;t]\right\}+o(\alpha^4)\, ,
\en
where
\eq\label{R}
{\cal R}[b;t]&=&\int\limits_{0}^{\infty}\frac{dpdk}{pk}\ln\frac{(p-k)^2+b^2}
{(p+k)^2+b^2}\,t(2M;0,p)t(2M;k,0)\\[3mm]\nonumber
&+&
\int\limits_{0}^{\infty}\frac{dpdk}{pk}\left\{\ln\frac{(p-k)^2}{(p+k)^2}-
\ln\frac{(p-k)^2+b^2}{(p+k)^2+b^2}\right\}\left [t(2M;0,p)t(2M;k,0)-
\bar{t}^{\,\, 2} 
\right ]
\en

It can be checked that the result does not depend on the arbitrary cutoff
parameter $b$.

{\bf 3.} The integration in the matrix element containing many-Coulomb 
Green function, can be directly carried out. The result is
\eq\label{n-C}
(\Psi_0|t(E_0)\,\bar{\bf g}^{n-C}(E_0)\,t(E_0)|\Psi_0) = 
-\frac{3\alpha M^2}{8\pi}\,\phi_0^2\, \bar{t}^{\,\, 2}+o(\alpha^4)\, .
\en

Finally, putting things together, for the energy shift we obtain

\eq
\Delta E=\phi_0^2\left\{\bar{t}-\frac{\alpha M^2}{4\pi}\,\bar{t}^{\,\, 2}-
\frac{\alpha M^2}{4\pi}\,\bar{t}^{\,\, 2}\,\ln\frac{b}{2\gamma}\,+
\frac{4\alpha M}{\pi}\,{\cal Q}[t]+\frac{\alpha M^2}{4\pi^3}{\cal R}[b;t]\right\}
+o(\alpha^4)\, .
\en

\noindent
The functionals ${\cal R}[b;t]$ and ${\cal Q}[t]$ are given above. The expression
for the energy shift can be rewritten in form similar to Eq.~(\ref{energyshift})
\eq\label{E-NR}
\Delta E=-\frac{\alpha^3 M^3}{8\pi}\,{\cal A}^{NR}\left\{1-
\frac{\alpha M^2}{4\pi}\,{\cal A}^{NR}(\ln\alpha-1)\right \}+o(\alpha^4)\, ,
\en
where
\eq\label{A-NR}
{\cal A}^{NR}=-\,\bar{t}+\frac{\alpha M^2}{8\pi}\,\bar{t}^{\,\, 2}\,\ln\frac{b^2}{M^2}-
\frac{4\alpha M}{\pi}\,{\cal Q}[t]-\frac{\alpha M^2}{4\pi^3}\,
{\cal R}[b;t]+o(\alpha)\, .
\en

In order to check the universality, one has to evaluate the full scattering
matrix $T$ which is defined by Eq.~(\ref{LS}), in the vicinity of threshold.
We use the scattering theory on two potentials, and evaluate $T$ 
perturbatively at $O(\alpha)$
\eq\label{sc2}
T={\bf h}_C+t+{\bf h}_C{\bf g}_0t+t{\bf g}_0{\bf h}_C+t{\bf g}_0{\bf h}_C{\bf
  g}_0
t+o(\alpha)\, ,\quad\quad \hat T=T-{\bf h}_C\, .
\en

The integrals entering here, are calculated similar to given above.
Note that, even there are no more ultraviolet divergences, we still use
dimensional regularization in order to regularize infrared divergences. 
After removal the Coulomb phase, one may safely put
$d\rightarrow 3$. The truncated amplitude in the vicinity of threshold 
behaves as
\begin{equation}
{\rm Re}\,[e^{-2i\alpha\theta_c}({\bf q}|{\tilde T}|{\bf p})]
\biggr|_{E={\bf q}^2/M,~|{\bf q}|=|{\bf p}|\rightarrow 0}=
\frac{\alpha\pi M}{2 |{\bf p}|}\,\bar{t}+
\frac{\alpha M^2}{4\pi}\,\bar{t}^{\,\, 2}\,\ln\frac{2|{\bf p}|}{M}-
{\cal A}^{NR}+O(|{\bf p}|)\, ,
\end{equation}
with exactly the same ${\cal A}^{NR}$ as in Eq.~(\ref{A-NR}). This means 
that, we have verified the universality conjecture formulated in the 
beginning of this section.

\subsection{Matching condition for the potential}

The universality conjecture, proven in the previous section for the
one-channel case, provides one with the matching condition for the short-range
potential $u({\bf q},{\bf p})$. In order to derive this condition, we assume
that short-range potential, in analogy with Eq.~(\ref{exp-1c}), can be written
as
\eq\label{u0u1}
u({\bf q},{\bf p})=u_0({\bf q},{\bf p})
+\alpha u_1({\bf q},{\bf p})+O(\alpha^2)\, . 
\en
Further, we introduce
\eq
t_0=u_0+u_0\,{\bf g}_0\,t_0\, .
\en
Using scattering theory on two potentials, we obtain
\eq
t=t_0+\alpha\,(1+t_0{\bf g}_0)\,u_1\,(1+{\bf g}_0t_0)+O(\alpha^2)
=t_0+\alpha t_1+O(\alpha^2)\, .
\en
The matching condition for the potential can be obtained by using
Eqs.~(\ref{regular-1c}), (\ref{exp-1c}) and (\ref{A-NR}) order by order in 
$\alpha$
\eq\label{general-1c}
{\cal A}_0^R&=&-4M^2\,\bar{t}_0\, ,
\\[2mm]
{\cal A}_1^R&=&-4M^2\,\biggl\{ (1+t_0{\bf g}_0)u_1(1+{\bf g}_0t_0)
-\frac{M^2}{8\pi}\,\bar{t}_0^{\,\, 2}\ln\frac{b^2}{M^2}
+\frac{4M}{\pi}\,{\cal Q}[t_0]
+\frac{M^2}{4\pi^3}\, {\cal R} [b;t_0] \biggr\}\, ,
\nonumber
\en
where $\bar{t}_0={\rm Re}\, t_0(2M;0;0)$.

It is seen, that the matching condition imposes rather loose constraints on
the potential $u({\bf q},{\bf p})$. For example, the matching condition  
in the first line requires that the strong scattering lengths in the
relativistic theory and in the potential model are the same. The behavior of
the scattering matrix above threshold does not play any role. This property of
the matching condition is not surprising, if one adopts the interpretation of
the potential model given in the previous section, namely that the short-range
potentials are regularizations of the contact interactions that arise from
field theory. The looseness of the matching condition then merely reflects the
freedom in the choice of such a regularization. Using this freedom, one may
further specify the potential, assuming
\eq
u_1({\bf q},{\bf p})=\lambda u_0({\bf q},{\bf p})\, .
\en
This amounts to the matching of the ``strength'' of the potential, that is the
counterpart of the coupling in the non-relativistic Lagrangian, at
$O(\alpha)$, with the momentum dependence of the potential fixed by hand. Using the
above ansatz, one can rewrite the matching condition in the following form
\eq\label{special-1c}
{\cal A}_0^R&=&-4M^2\,\bar{t}_0\, ,
\nonumber\\[2mm]
{\cal A}_1^R&=&-4M^2\,\biggl\{ \lambda\,\biggl(\bar{t}_0
-\frac{M}{2\pi^2}\,{\cal G}[t_0]\biggr)
-\frac{M^2}{8\pi}\,\bar{t}_0^{\,\, 2}\ln\frac{b^2}{M^2}
+\frac{4M}{\pi}\,{\cal Q}[t_0]
+\frac{M^2}{4\pi^3}\, {\cal R} [b;t_0] \biggr\}\, ,
\en
where
\eq
{\cal G}[t]=\int_0^\infty dp\, t(2M;0,p)\, t(2M;p,0)\, .
\en

This matching condition uniquely determines the couplings in the potential
at $O(1)$ and $O(\alpha)$. Namely, first, one chooses the momentum dependence
of the potential $u_0({\bf q},{\bf p})$ and adjusts the coupling at $O(1)$,
to reproduce the strong scattering length obtained from the relativistic
theory (the first line of Eq.~(\ref{special-1c})). This completely determines
the scattering matrix $t_0$ at all momenta and energies. At the next step,
with a given $t_0$, one calculates all integrals entering the matching
condition at $O(\alpha)$ (second line of Eq.~(\ref{special-1c})), and
determines the coupling constant $\lambda$ from this equation.
Despite the unrestricted freedom in the choice of the momentum dependence of
$u_0({\bf q},{\bf p})$, the potential $u({\bf q},{\bf p})$ that we construct,
reproduces, by construction, the strong energy-level shift at $O(\alpha^4)$,
as well as threshold scattering amplitude at $O(\alpha)$.

\subsection{Separable potential}

For demonstrative purposes, below we again consider the solution of matching
equation for simple rank-1 separable potential
\eq\label{sep}
u_0=g\,u({\bf q})\,u({\bf p}); \quad\quad 
u({\bf q})=\frac{\beta^2}{\beta^2+{\bf q}^2}
\en

With this potential, one is able to carry out the calculation of all
quantities entering the matching condition, in a closed form.
The result is given by
\eq
&&{\bar t}_0=g\,\left(1+g\,\frac{\beta M}{8 \pi}\right )^{-1}\, ,\quad\quad
{\cal Q}[t_0]= -\frac{\pi}{2\beta}\,{\bar t}_0\, ,\nonumber\\[2mm]
&&{\cal R}[b;t_0]=\frac{\pi^2}{2}\,{\bar t}_0^{\,\, 2}\ln\frac{4 b^2}{\beta^2}\, ,
\quad\quad ~
{\cal G}[t_0]=\frac{\pi}{4}\,\beta\, {\bar t}_0^{\,\, 2}
\en

Substituting these expressions into the Eq.~(\ref{special-1c}), one 
obtains two equations which fix the couplings $g$ and $\lambda$ at $O(1)$ and
$O(\alpha)$, respectively
\eq\label{match-sep}
g&=&-\frac{{\cal A}_0^R}{4M^2\left(1+\frac{\beta{\cal A}_0^R}{32\pi M}\right)}\, ,
\nonumber\\[2mm]
\lambda&=&\frac{1}{1+\frac{\beta{\cal A}_0^R}{32\pi M}}
\biggl(\frac{{\cal A}_1^R}{{\cal A}_0^R}+\frac{2M}{\beta}
-\frac{1}{32\pi}\,{\cal A}_0^R\,\ln\,\frac{\beta^2}{4M^2}\biggr)\, .
\en
As expected, the potential range parameter $\beta$
is not determined from matching condition.

We mention again, that the separable form of the strong potential~(\ref{sep}) is not, by far, the
unique choice. E.g., the local square-well strong potentials used in
Ref.~\cite{Rasche}, can serve equally well. In the latter case, however,
we were not able to obtain the corresponding integrals that enter the matching
condition, in a closed form. This means, that numerical methods should be 
used.

\section{Two-channel case: decay of the pionium}
\setcounter{equation}{0}
\label{sec:two-channel}

In the previous Section we have considered the strong energy shift of the 
hadronic bound state at $O(\alpha^4)$. The problem was solved in the
one-channel model - that is, the theory contained, from the beginning, only a
doublet of charged scalar particles. In addition, we have neglected
relativistic corrections and higher-order derivative couplings in the strong
Lagrangian. As was mentioned, all these effects do not contribute to the
strong energy shift at the accuracy considered, and the (possible) coupling 
to
other channels does not show up explicitly in the energy-shift calculations 
at 
$O(\alpha^4)$ as well~\cite{Bern3}. The situation is quite different, if one considers the
decay of the hadronic bound state at next-to-leading order in isospin
breaking: here all above effects should be consistently taken into account. 
In this Section, we shall enlarge the formalism developed in the
Section~\ref{sec:one-channel} to the two-channel case, and include relativistic
effects as well as derivative interactions. To this end, we consider the
decay of the $\pi^+\pi^-$ atom (pionium) which has been already investigated in detail 
within the non-relativistic effective Lagrangian approach and
ChPT~\cite{Bern1,Bern2,Bern4}. Generalization to other hadronic atoms,
as well as inclusion of spin effects is straightforward an will not be
discussed.

Preliminary remarks are in order. The isospin breaking in the $\pi^+\pi^-$
system is due to two physically distinct sources: electromagnetic corrections
which are parameterized by the fine structure constant $\alpha$, and the
quark  mass
difference $m_d-m_u$. The notions of ``isospin-symmetric world'' and ``pure
strong interactions'' refer to the idealized world with $\alpha=0$,
$m_d-m_u=0$, and where, by convention, the mass of the pion coincides with 
the
charged pion mass in the real world. It is convenient to introduce the common
counting for two different isospin breaking parameters. In the following, we
shall use the following assignment~\cite{Bern1,Bern2,Bern4}
\eq
\alpha\sim(m_d-m_u)^2\sim\delta\, .
\en

The $\pi^+\pi^-$ atom decays predominantly into the $\pi^0\pi^0$ final state:
$\Gamma_{tot}=\Gamma_{2\pi^0}+O(\delta^5)$~\cite{Bern1,Bern2,Bern4}.
At leading order, $\Gamma_{2\pi^0}=O(\delta^{7/2})$. We shall be interested
in next-to-leading order corrections in isospin breaking - up to and including
order $\delta^{9/2}$. At this accuracy, still only the decay into $\pi^0\pi^0$
final state is possible~\cite{Bern4}, so it is perfectly consistent to
restrict ourselves to the consideration of the two-channel ($\pi^+\pi^-$ and
$\pi^0\pi^0$) problem in the quantum-mechanical framework. 

Our strategy will be follows. First, we consider the ``relativization'' of the
non-relativistic effective Lagrangian approach used in
Refs.~\cite{Bern1,Bern2,Bern4}, in order to bring this approach in
conformity with the relativized potential model used in Refs.~\cite{Rasche,Rasche-new,Rasche-talk}
to treat the same problem. The relativized field-theoretical framework
is used to calculate the decay width of the $\pi^+\pi^-$ atom at order
$\delta^{9/2}$ - the result, of course, agrees with that from
Ref.~\cite{Bern1}. Next, we construct the two-channel potential model which
reproduces the result of the field-theoretical approach. The universality
conjecture is verified for the two-channel potential model, with relativistic
corrections and derivative interactions taken into account. This provides us
with the matching conditions (one per channel) for the short-range potential,
which can be solved similar to the one-channel case.

\subsection{Relativistic corrections}

In Ref.~\cite{Bern4} it has been argued that the following non-relativistic
effective Lagrangian is sufficient to carry out the calculation of the decay
width of the $\pi^+\pi^-$ atom at $O(\delta^{9/2})$
\eq\label{Lagrangian-pi}
{\cal L}&=&{\cal L}_0 + {\cal L}_D + {\cal L}_C + {\cal L}_S \nonumber\\ [2mm]
{\cal L}_0&=& \sum_{i=\pm, 0}\,\pi_i^\dagger\biggl( i\partial_t-M_{\pi^i}+
\frac{\triangle}{2M_{\pi^i}}\biggr)\pi_i,\nonumber\\[2mm]
{\cal L}_D&=&\sum_{i=\pm, 0}\,\pi_i^\dagger
\biggl(\frac{\triangle^2}{8M_{\pi^i}^3}+\cdots\biggr)\pi_i,\\
{\cal L}_C&=&-4\pi\alpha(\pi_-^\dagger\pi_-)\triangle^{-1}(\pi_+
^\dagger\pi_+)+\cdots ,\nonumber\\[2mm]
{\cal L}_S&=&c_1\pi_+^\dagger\pi_-^\dagger\pi_+\pi_-
+c_2[\pi_+^\dagger\pi_-^\dagger(\pi_0)^2+{\rm h.c.}]
+c_3\,(\pi_0^\dagger\pi_0)^2\nonumber\\[2mm]
&+&c_4[\pi_+^\dagger\stackrel{\leftrightarrow}
{\triangle}\pi_-^\dagger(\pi_0)^2+\pi_+^\dagger\pi_-^\dagger\pi_0
\stackrel{\leftrightarrow}{\triangle}\pi_0+{\rm h.c.}]+\cdots\, .\nonumber
\en
This Lagrangian is built from the non-relativistic pion field 
$\pi_{\pm},\pi_0$. It contains the non-relativistic kinetic term
${\cal L}_0$, along with the term
${\cal L}_D$ which accounts for the relativistic corrections to the pion
energy. These corrections have been included in the bound-state width in a
perturbative manner~\cite{Bern1,Bern4}.
Further, the formation of the bound state proceeds mainly due to the static
Coulomb interaction contained in ${\cal L}_C$, whereas strong interactions
described by the local four-pion Lagrangian ${\cal L}_S$ are mainly
responsible for its decay. The constants in ${\cal L}_S$ are determined from
matching to the relativistic theory. Again, as in the one-channel case, we have
truncated all terms which do not contribute to the quantity of interest (the
decay width) at the accuracy we are working.

In order to bring our framework in conformity with the relativized potential 
model which have been used for the study of pionium decay~\cite{Rasche,Rasche-new,Rasche-talk}, 
below we include the relativistic corrections contained in ${\cal L}_D$, in
the unperturbed Lagrangian. Diagrammatically, this corresponds to summing up
all mass insertions in the free non-relativistic pion propagator (see
Fig.~\ref{fig:sum}). In actual calculations of the diagrams in effective field
theory, one is 
again forced to treat part of these corrections perturbatively, in the
expansion of powers of ${\bf p}^2/M_{\pi^i}^2$. Since our trick amounts merely to the
redistribution of terms in the total Lagrangian between unperturbed and
interaction pieces, the results for any observable quantity (e.g. the
decay width) should remain unaffected. The reason why this redistribution is
carried out, is twofold.
\begin{itemize}
\item[i)]{If one has the same bound-state equations in the effective field
    theory and in the potential model, one can merely read off the
    potential from field-theoretical bound-state equations.}
\item[ii)]{Though the perturbative treatment of the mass insertions in the
    effective field theory is easy, this becomes rather complicated in the
    potential model with general non-contact interactions. Technically, it is
    preferable to have a framework where these insertions are summed up from
    the beginning.}
\end{itemize}

In order to design such a framework, we bring together ${\cal L}_0$ and
${\cal L}_D$ to form the relativized kinetic term
\eq
{\cal L}_R={\cal L}_0+{\cal L}_D=\sum_{i=\pm,0}\,
\pi_i^{\dagger}\,\left (i\partial_t-\sqrt{M_{\pi^i}^2-\triangle}\,
\right )\,\pi_i\, .
\en
The corresponding free relativized Hamiltonian is given by
\eq
{\bf H}_R=\int d^3{\bf x}\, \sum_{i=\pm,0}\pi_i^\dagger(0,{\bf x})\,
\sqrt{M^2-\triangle}\,\,\,\pi_i(0,{\bf x})\, ,
\en
and Coulomb (${\bf H}_C$) and strong (${\bf H}_S$) Hamiltonians are defined in
analogy with Eq.~(\ref{Hamiltonian}).

The full scattering matrix obeys the Lippmann-Schwinger equation
\eq
T'(z)=({\bf H}_C+{\bf H}_S)+({\bf H}_C+{\bf H}_S)\,{\bf G}'(z)\,T'(z)\, ,\quad\quad
{\bf G}'(z)=\frac{1}{z-{\bf H}_R}\, .
\en

Poles of the scattering matrix on the second Riemann sheet of the complex
$z$-plane correspond to unstable bound states. The real and imaginary parts of
the pole position determine the energy and width of such a bound state,
according to $E={\rm Re}\, z$, $\Gamma=-2{\rm Im}\, z$.

Further, we define the transformed quantities
\eq\label{transformation}
~_A\langle {\bf Q},{\bf q}|\tilde T (z)|{\bf P},{\bf p}\rangle_B&=&
~_A\langle {\bf Q},{\bf q}|\left (\frac{z+{\bf H}_R}{4 M_As_A}\right )^{1/2}T'(z)\left
  (\frac{z+{\bf H}_R}{4 M_Bs_B}\right )^{1/2}|{\bf P},{\bf p}\rangle_B \nonumber\\[2mm]
~_A\langle {\bf Q},{\bf q}|\tilde V (z)|{\bf P},{\bf p}\rangle_B&=&
~_A\langle {\bf Q},{\bf q}|\left (\frac{z+{\bf H}_R}{4 M_As_A}\right )^{1/2}
({\bf H}_C+{\bf H}_S)
\left(\frac{z+{\bf H}_R}{4 M_Bs_B}\right )^{1/2}|{\bf P},{\bf p}\rangle_B
\\[2mm]
~_A\langle {\bf Q},{\bf q}|\tilde {\bf G}_0 (z)|{\bf P},{\bf p}\rangle_B&=&
~_A\langle {\bf Q},{\bf q}|\left( \frac{4M_A}{s_A(z+{\bf H}_R)}\right)^{1/2}
{\bf G}'(z)
\left( \frac{4M_B}{s_B(z+{\bf H}_R)}\right)^{1/2}|{\bf P},{\bf p}\rangle_B\, .
\nonumber
\en
with $s_+=1,~s_0=2!$ and $A,B=+,0$ label the charged ($\pi^+\pi^-$) and 
neutral
($\pi^0\pi^0$) channels.

In the CM frame, the free resolvent (see Eq.~(\ref{transformation})) takes the
form
\eq\label{Rasche-GF}
~_A({\bf q}|\tilde{\bf g}_0(z)|{\bf p})_B
=(2\pi)^3\delta^3({\bf q}-{\bf p})\,
\pmatrix{\frac{\displaystyle{M_{\pi^+}}}{\displaystyle{k_+^2-{\bf p}^2}}&0\cr
0&\frac{\displaystyle{M_{\pi^0}}}{\displaystyle{k_0^2-{\bf p}^2}}}_{AB}\, ,
\en
where
\eq\label{q+0}
k_+^2=\frac{z^2}{4}-M_{\pi^+}^2\, ,\quad\quad
k_0^2=\frac{z^2}{4}-M_{\pi^0}^2\, .
\en
We see that, despite the non-relativistic appearance of the free
resolvent~(\ref{Rasche-GF}), the relativistic effects are taken into account
in the definition of the quantities $k_+^2,~k_0^2$, Eq.~(\ref{q+0}).

The Lippmann-Schwinger equation for the transformed scattering matrix now
exactly corresponds to the Schr\"{o}dinger equation used in the relativized
potential model~\cite{Rasche}
\eq\label{LS-R}
\tilde t(z)=\tilde v(z)+\tilde v(z)\tilde{\bf g}_0(z)\tilde t(z)\, .
\en

If one expands the transformed potential in Eq.~(\ref{transformation}) in
the small quantities ${\bf p}^2/M_{\pi^i}^2$, $z-2M_{\pi^+}\sim O(\delta^2)$
and $z-2M_{\pi^0}\sim O(\delta)$, it is seen that the transformation formally 
amounts to the following replacements
\eq
c_1&\rightarrow&c_1\left\{1+\frac{z-2 M_{\pi^+}}{4
    M_{\pi^+}}\right\}+\cdots\, ,\nonumber\\[2mm]
c_2&\rightarrow&c_2\left\{1+\frac{z-2 M_{\pi^+}}{8
    M_{\pi^+}}+\frac{z-2 M_{\pi^0}}{8M_{\pi^0}}\right \}+\cdots\, ,\nonumber\\[2mm]
c_3&\rightarrow&c_3\left\{1+\frac{z-2 M_{\pi^0}}{4
    M_{\pi^0}}\right\}+\cdots\, ,\nonumber\\[2mm]
c_4&\rightarrow&c_4-\frac{c_2}{16M_{\pi^+}^2}+ \cdots\, .
\en

Since the matching condition determines these couplings from the amplitudes
evaluated at threshold, one has to substitute $z=2M_{\pi^+}$ or $z=2M_{\pi^0}$
(depending on the particular channel) in the above expressions. Finally, one
may conclude that the transformation, introduced above, amounts to the
following redefinition of strong couplings in the
Lagrangian~(\ref{Lagrangian-pi}) 
\eq\label{primes}
&&
c_1'=c_1+\cdots\, ,\quad
c_2'=c_2\,\biggl( 1+\frac{\Delta M_\pi^2}{8M_{\pi^+}^2}\biggr)+\cdots\, ,\quad
c_3'=c_3+\cdots\, ,\quad
c_4'=c_4-\frac{c_2}{16M_{\pi^+}^2}+ \cdots\, .
\nonumber\\
&&
\en

Accordingly, one may define the energy-independent potential $\bar V$ 
which is obtained from the transformed potential $\tilde V(z)$ by 
substituting the threshold values for the parameter $z$.
In the CM frame this potential is given by
\eq\label{v-tr}
({\bf q}|\bar{v}|{\bf p})=-\frac{4\pi\alpha}{|{\bf q}-{\bf p}|^2}+
\pmatrix{-c_1'& \sqrt{2}(-c_2'+2c_4'({\bf q}^2+{\bf p}^2))\cr
&\cr
\sqrt{2}(-c_2'+2c_4'({\bf q}^2+{\bf p}^2))& -2c_3'}+\cdots\, .
\en
By construction, the energy-independent potential $\bar v$, when used in the
Lippmann-Schwinger equation (\ref{LS-R}), reproduces the decay width in the
first non-leading order in isospin breaking.

Note that the relativization scheme described above is linked to
the particular choice of the Lippmann-Schwinger equation in
Ref.~\cite{Rasche}. 
If a different choice is assumed, it is straightforward to adapt the above
relativization scheme to this new equation.

\subsection{Decay width of the pionium} 

One may study the pionium decay in the relativized framework by using exactly
the same methods as within the non-relativistic effective Lagrangian approach.
We refer reader to Refs.~\cite{Bern1,Bern2,Bern4} for more details concerning
the technique, and merely quote the final results here. The decay width is
given by
\eq\label{width}
\Gamma_{2\pi^0}=\frac{2}{9}\,\alpha^3p^{\star}{\cal A}^2\, (1+K)\, .
\en
where
$p^{\star}=(M_{\pi^+}^2-M_{\pi^0}^2-\frac{1}{4}\,\alpha^2M_{\pi^+}^2)^{1/2}$ 
and the quantities ${\cal A}$ and $K$ are defined below.

The matching condition now reads
\eq
~_A({\bf q}|\tilde t(z)|{\bf p})_B
=-\,\frac{1}{2\sqrt{M_As_A\,\omega_A({\bf q})}}\,
T_{AB}^{R}({\bf q},{\bf p})\frac{1}{2\sqrt{M_Bs_B\,\omega_B({\bf p})}}\,
,\quad\quad A,B=+,0\, .
\en

The threshold behavior of the relativistic $\pi^+\pi^-\rightarrow\pi^0\pi^0$
scattering amplitude at $O(\delta)$ is given by (cf with Eq.~(\ref{R-1c-thr})) 
\eq
{\rm Re}\,(e^{-i\alpha\theta_C}T^R_{+0}({\bf q},{\bf p}))
\biggr|_{|{\bf q}|\rightarrow 0}=
\frac{\tilde B_1^R}{|{\bf q}|}+\tilde B_2^R\,\ln\frac{2|{\bf q}|}{M_{\pi^+}}
+{\rm Re}\, A_{\rm thr}^{+-00}+O(|{\bf q}|)\, .
\en
And the quantity ${\cal A}$ is, from the matching condition, expressed through
the regular part of the relativistic $\pi^+\pi^-$ scattering amplitude at
threshold
\eq
{\cal A}=-\frac{3}{32\pi}\,{\rm Re}\,A_{\rm thr}^{+-00}+o(\delta)\, ,\quad\quad
{\cal A}=a_0-a_2+O(\delta)\, ,
\en
where $a_I$ denote the $S$-wave $\pi\pi$ scattering lengths in the isospin
limit in the channel with total isospin $I$. Further, the quantity $K$ is
expressed through these scattering lengths according to
\eq\label{K-pipi}
K&=&\frac{\kappa}{9}\,(a_0+2a_2)^2-\frac{2\alpha}{3}\,(\ln\alpha-1)
(2a_0+a_2)+o(\delta)\, .
\en
with $\kappa=M_{\pi^+}^2/M_{\pi^0}^2-1$.

The matching condition for the couplings $c_1'\cdots c_4'$ is given by
\eq
\frac{1}{4M_{\pi^+}^{3/2}M_{\pi^0}^{1/2}}\, {\rm Re}\, A^{+-00}_{\rm thr}&=&
2c'_2-4M_{\pi^0}^2\kappa\biggl( c'_4+\frac{c'_2{c'_3}^2
M_{\pi^0}^2}{8\pi^2}\biggr)
+\frac{\alpha M_{\pi^+}^2}{4\pi}\,\biggl( 1-\Lambda(\mu) 
-\ln\frac{M_{\pi^+}^2}{\mu^2}\biggr)\, c'_1c'_2\, ,
\nonumber\\[2mm]
c'_1&=&\frac{4\pi}{3M_{\pi^+}^2}\,(2a_0+a_2)\, ,
\nonumber\\[2mm]
c'_3&=&\frac{2\pi}{3M_{\pi^+}^2}\,(a_0+2a_2)\, .
\en

This matching condition determines $c'_1,c'_2,c'_3$ at $O(1)$. In addition,
it fixes a particular linear combination of $c'_2$ at
$O(\delta)$ and $c'_4$ at $O(1)$. Having expressed these couplings through the
relativistic scattering amplitude, one may again interpret the
short-range part of the energy-independent potential~(\ref{v-tr}) 
as the contact potential to be used in the Schr\"{o}dinger equation - provided
one uses the dimensional regularization throughout. However, in a complete
analogy with the one-channel case, there is also a possibility to construct the
conventional short-range potential which reproduces the field-theoretical
result for the width at $O(\delta^{9/2})$ and does not lead to the ultraviolet
divergence. This is demonstrated in the following Section.

\subsection{Pionium decay within the potential model}

In this Section, we shall demonstrate that universality, which we have 
proven for
the ground-state energy-level shift in the one-channel case, also holds for
the decay width of the pionium. Namely, we prove that {\em the relation
between the decay width of the ground state and the threshold amplitudes is
the same in field theory and in the potential model in the first non-leading
order in isospin breaking.} Given the fact, that the energy and the decay
width are the only observable characteristics of the bound state, one may
conclude that the properties of the bound state in the first non-leading
order in isospin breaking are completely determined by the properties of the
amplitudes. 

The calculations which are performed below, are analogous to the ones already
carried out in the one-channel case (see Section~\ref{sec:one-channel}). We give only
the final results here.

The scattering matrix in the CM frame obeys the Lippmann-Schwinger equation
\eq\label{Rasche}
\tilde t(z)=v+v\,{\tilde {\bf g}}_0(z)\,{\tilde t}(z)\, .
\en
All quantities entering this equation are $2\times 2$ matrices with the
entries $A,B=+,0$. In particular, the potential of the above
Lippmann-Schwinger equation is given by
\eq\label{pot_full}
~_A({\bf q}|v|{\bf p})_B=-\frac{4\pi\alpha}{|{\bf q}-{\bf p}|^2}\,
\delta^{A+}\delta^{B+}+u^{AB}({\bf q},{\bf p})\doteq
({\bf q}|{\bf h}_C|{\bf p})\,\delta^{A+}\,\delta^{B+}
+u^{AB}({\bf q},{\bf p})\, ,
\en
where $u^{AB}({\bf q},{\bf p})$ stands for the short-range part of the
potential. 

The two-channel counterpart of the scattering matrix $\tau(z)$ introduced in
Eq.~(\ref{tau-f}) satisfies the following equation
\eq
\tau (z)=u+u\,\tilde {\bf g}_C(z)\,\tau (z)\, ,
\en
where the pole removed Coulomb Green function is given by
\eq
~_A({\bf q}|\tilde {\bf g}_C(z)|{\bf p})_B=\pmatrix{
({\bf q}|\bar{\bf g}_C(2M_{\pi^+}+k_+^2(z)/M_{\pi^+})|{\bf p})&0\cr
&\cr
0&({\bf q}|{\bf g}_0(2M_{\pi^0}+k_0^2(z)/M_{\pi^0})|{\bf} p)}\, .
\en

Further, in analogy with Eq.~(\ref{def-t}), we again introduce
\eq
t(z)=u +u\,\tilde{\bf  g}_0(z)\,t(z)\, ,\quad
\tau(z)=t(z)+t(z)\,\delta \tilde{\bf g}(z)\,\tau (z)\, ,\quad
\delta \tilde{\bf g}(z)=\tilde {\bf g}_C(z)-\tilde {\bf g}_0(z)\, .
\en

We again, as in the one-channel case, assume, that the potential $u^{AB}$
is hermitian and real. In addition, we assume that $u^{AB}=u^{BA}$.
The unitarity condition for the scattering matrix $t(z)$ in the two-channel
case takes the form
\eq
{\rm Im}\, t^{AB}(z;q,p)&=&-\frac{M_{\pi^+} k_+(z)}{4\pi}\,
t^{A+}(z;q,k_+(z))\,(t^{+B}(z;k_+(z),p))^{\star}\nonumber\\[2mm]
&-&\frac{M_{\pi^0} k_0(z)}{4\pi}\,
t^{A0}(z;q,k_0(z))\,(t^{0B}(z;k_0(z),p))^{\star}\, ,
\en
and the partial-wave expansion is again performed, according to
formulae~(\ref{partial}).

\subsubsection{Decay width}

The position of the bound-state pole in the full scattering matrix is
determined by the equation (cf with Eq.~(\ref{Pole_position}))
\eq\label{potential-pole}
z-E_0-(\Psi_0|\,\tau^{++}(z)\,|\Psi_0)=0\, .
\en

Solving this equation by iterations, it is straightforward to
demonstrate (see Appendix~\ref{app:decay}) that the decay width
$\Gamma=-2{\rm Im}\, z$ can be rewritten in a form~(\ref{width}), if we
identify
\eq\label{A-decay}
{\cal A}&=&\frac{3}{32\pi}\,4\sqrt{2}\,M_{\pi^+}^{3/2}M_{\pi^0}^{1/2}\left\{ 
{\rm Re}\, {\bar t}^{\,\,+0}-\frac{\alpha M_{\pi^+}^2}{4\pi}\,
\ln\frac{b}{M_{\pi^+}}\,{\rm Re}\, {\bar t}^{\,\,+0}
{\rm Re}\, {\bar t}^{\,\,++}\right.
\nonumber\\[3mm]
&+&\left. \frac{2\alpha M_{\pi^+}}{\pi}\,{\tilde {\cal Q}}[t]
+\frac{\alpha M_{\pi^+}^2}{4\pi^3}\, {\tilde {\cal R}}[b;t]\right\}+
O(\delta^2)\, ,
\en
where the following notations are introduced
\eq
{\bar t}^{++}=t^{++}(z_+;0,0)\, ,\quad\quad
{\bar t}^{+0}=t^{+0}(z_+;0,k_0(z_+))\, ,\quad\quad
{\bar t}^{00}=t^{00}(z_0;0,0)\, ,
\en
and $z_+=2M_{\pi^+}$ and $z_0=2M_{\pi^0}$. The functionals
$\tilde{\cal Q}[t]$ and $\tilde{\cal R}[b;t]$ are defined in
Appendix~\ref{app:decay}.

\subsubsection{Matching condition for the threshold
  scattering amplitudes}

In order to check the universality conjecture, we calculate the scattering
amplitude for the process $\pi^+\pi^-\rightarrow\pi^0\pi^0$ in the vicinity of
threshold, in the potential model. The perturbative expansion in $\alpha$ for 
the scattering matrix has the form
\eq
{\tilde t}^{\,\,+0}=t^{+0}+{\bf h}_C\, ({\tilde {\bf g}}_0)^{++}\, t^{+0} + 
t^{++}\, ({\tilde {\bf g}}_0)^{++}\, {\bf h}_C\,
({\tilde {\bf g}}_0)^{++}\, t^{+0}+ o(\delta)\, . 
\en
In order to calculate the integrals entering here, one has to use the
dimensional regularization to regulate the infrared divergence caused by the
one Coulomb photon exchange. At threshold, the behavior of the scattering 
matrix is given by
\eq\label{threshold-pot2}
{\rm Re}\,\bigl[e^{-i\alpha\theta_C}~_+({\bf q}|\tilde t|{\bf p})_0\bigr]
\biggr|_{|{\bf q}|\to 0\, ,\,|{\bf p}|\to k_0(z_+)} &=& 
-\frac{\alpha\pi M_{\pi^+}}{4|{\bf q}|}\, {\bar t}^{\,\,+0}-
\frac{\alpha M_{\pi^+}^2}{4\pi}\,{\rm Re}\, {\bar t}^{\,\,++}\,
{\rm Re}\, {\bar t}^{\,\,+0}\,\ln\frac{2|{\bf q}|}{M_{\pi^+}}
\nonumber\\[2mm]
&+&
\frac{1}{4\sqrt{2}\,M_{\pi^+}^{3/2} M_{\pi^0}^{1/2}}\,
\frac{32\pi}{3}\, {\cal A}+O(|{\bf q}|)\, .
\en

In addition, we shall use the matching condition in the isospin limit in the
elastic channels $\pi^+\pi^-\rightarrow\pi^+\pi^-$ and 
$\pi^0\pi^0\rightarrow\pi^0\pi^0$
\eq
{\rm Re}\, {\bar t}^{\,\,++}=-\frac{4\pi}{3 M_{\pi^+}^2}\, 
(a_2+2 a_0)+O(\delta)\,
,
\quad\quad
{\rm Re}\, {\bar t}^{\,\,00}=-\frac{4\pi}{3 M_{\pi^+}^2}\, (a_0+2 a_2)+O(\delta)\,
.
\en

One may straightforwardly check, that the regular term ${\cal A}$ in
Eq.~(\ref{threshold-pot2}) is
indeed given by the expression~(\ref{A-decay}) obtained from bound-state
calculations. Consequently, the universality conjecture holds also in the
two-channel model, in the context of the bound-state decay problem.

\subsection{Matching procedure for the potential}

Using the explicit expression~(\ref{A-decay}), we shall derive the matching
condition for the short-range potential by matching the amplitudes at
threshold in the relativistic theory and in the potential model. The
universality then ensures, that the bound state decay width is the same in
both theories at $O(\delta^{9/2})$.  

As in the one-channel case, we assume that the short-range potential contains
the isospin-conserving and isospin-breaking parts
\eq
u^{AB}({\bf q},{\bf p})=u_0^{AB}({\bf q},{\bf p})+
u_1^{AB}({\bf q},{\bf p}) + o(\delta^2)\, ,
\quad\quad
u_0^{AB}\sim O(1)\, ,\quad\quad u_1^{AB}\sim O(\delta)\, ,
\en
and the matching condition holds separately at $O(1)$ and $O(\delta)$.

It is convenient to introduce the purely strong scattering matrix
\eq\label{IC}
t_0(z)=u_0+u_0\, {\bar {\bf g}}_0(z)\, t_0(z)\, ,
\en
where $\bar{\bf g}_0$ denotes the free resolvent in the absence of isospin
breaking 
\eq\label{Bar_g_0}
~_A({\bf q}|{\bar{\bf g}}_0(z)|{\bf p})_B
=(2\pi)^3\delta^3({\bf q}-{\bf p})\, \delta^{AB}\, 
\frac{M_{\pi^+}}{k_+^2(z)-{\bf p}^2}\, .
\en

The perturbative solution for the scattering matrix $t$ in terms of $t_0$ is
given by
\eq\label{split}
t=t_0+(1+t_0{\bar {\bf g}}_0)\, u_1\,(1+{\bar {\bf g}}_0 
t_0)+t_0 ({\tilde {\bf g}}_0-{\bar {\bf g}}_0) t_0
+ t_0 ({\tilde {\bf g}}_0-{\bar {\bf g}}_0) t_0 
({\tilde {\bf g}}_0-{\bar {\bf g}}_0) t_0+\cdots\, ,
\en
where ellipses stand for the terms that do not contribute to the real part of
the scattering matrix $t$ at $O(\delta)$. One has to mention here, that the
corrections in Eq.~(\ref{split}) containing 
${\tilde {\bf g}}_0-{\bar {\bf g}}_0$, are usually referred to as 
``mass-splitting corrections'' within the potential model. These corrections 
are,
of course, absent in the one-channel case. Moreover, it is implied in the
existing potential models, that these corrections fully account for the 
effects caused by the charged and neutral pion mass difference, that amounts
to the negligence of the mass corrections contained in the relativistic
amplitude.

On the other hand, the relativistic scattering amplitude that enters the
matching condition, can be written as~\cite{Bern2,Bern4}
\eq
{\cal A}=a_0-a_2+\epsilon+O(\delta^2)\, ,
\en
where, in order to calculate $\epsilon$ one invokes, e.g., ChPT.

Substituting these expressions into the matching condition for the amplitude
in the charge-exchange channel, we obtain
\eq\label{1delta}
&&\frac{32\pi}{3}\,\frac{1}{4\sqrt{2}\, M_{\pi^+}^{3/2}
  M_{\pi^0}^{1/2}}\,(a_0-a_2+\epsilon)=
\bar t_0^{+0}+\left\{(1+t_0 {\bar {\bf g}}_0)\,
u_1\,(1+{\bar{\bf g}}_0 t_0)\right\}^{+0}\biggr|_{\rm thr}
+\tilde{\cal F}[t_0]\, ,
\en
where $\tilde{\cal F}[t_0]$ is the complicated functional depending on
$t_0$. Its explicit form is given in Appendix~\ref{app:potential}.

The matching condition in the charge-exchange channel should be complemented
by two additional conditions from elastic channels, where it suffices to
perform matching at $O(1)$. Finally, at $O(1)$ the matching condition reads
\eq\label{m0}
\frac{4\sqrt{2}\,\pi}{3 M_{\pi^+}^2}\,(a_0-a_2)&=&\bar t_0^{\,\,+0}
\nonumber\\[2mm]
-\frac{4\pi}{3 M_{\pi^+}^2}\,(2a_0+a_2)&=&\bar t_0^{\,\,++}\nonumber\\[2mm]
-\frac{4\pi}{3 M_{\pi^+}^2}\,(a_0+2a_2)&=&\bar t_0^{\,\,00}
\en

At order $\delta$, from Eq.~(\ref{1delta}) we obtain the following matching
condition 
\eq\label{mdelta}
&&\frac{4\sqrt{2}\,\pi}{3 M_{\pi^+}^2}\,\left (\epsilon+
\frac{\kappa}{4}(a_0-a_2)\right )=\left\{(1+t_0\,{\bar{\bf g}}_0)\,u_1\,
(1+{\bar{\bf g}}_0\,t_0)\right\}^{+0}\biggr|_{\rm thr}
+\tilde{\cal F}[t_0]\, .
\en

As expected, the matching condition does not pose any constraint on the
momentum dependence of the short-range potential. We may use this freedom and
assume 
\eq\label{lambda_AB}
u_1^{AB}({\bf q},{\bf p})=\omega u_0^{AB}({\bf q},{\bf p})\, . 
\en

With this assumption, the integral entering the matching condition at
$O(\delta)$, can be expressed through $t_0$
\eq
\left\{(1+t_0{\bar {\bf g}}_0)u_1(1+{\bar {\bf g}}_0\, 
t_0)\right\}^{+0}\biggr |_{\rm thr}=\omega\left\{{\bar t}_0^{\,\,+0}-
\frac{M_{\pi^+}}{2\pi^2}{\tilde {\cal G}}[t_0]\right\}\, ,
\en
where
\eq
\tilde {\cal G}[t_0]=\int\limits_0^{\infty}dp\,\left\{t_0^{++}(z_+;0,p)\,
t_0^{+0}(z_+;p,0)+t_0^{+0}(z_+;0,p)\,t_0^{00}(z_+;p,0)\right\}\, ,
\en
and the matching condition then uniquely determines the coupling
$\omega$ in the short-range potential.

The matching procedure proceeds in several steps
\begin{itemize}
\item[i)]{One solves the equation~(\ref{IC}) in the absence of isospin 
breaking, in the basis of isospin scattering matrices $t_0,~t_2$, and adjusts
the couplings in the short-range potentials so, that the matching
conditions~(\ref{m0}) at $O(1)$ are satisfied. The
scattering matrices in physical channels are expressed via $t_0,~t_2$
according to    
\eq
t_0^{++}=\frac{1}{3}\,(2t_0+t_2)\, ,\quad\quad
t_0^{+0}=\frac{\sqrt{2}}{3}\,(t_2-t_0)\, ,\quad\quad
t_0^{00}=\frac{1}{3}\,(t_0+2t_2)\, .
\en
}
\item[ii)]{With the given $t_0^{++},~t_0^{+0},~t_0^{00}$ one
calculates the functional $\tilde{\cal F}[t_0]$
that appears in the right-hand side of the matching condition at
$O(\delta)$~(\ref{mdelta}). If the potential in each channel is given, e.g.
in the simple separable parameterization~(\ref{sep}), this again can be done
in a closed form. For the general short-range potential, one, however, may 
have to resort to the numerical methods. 
}
\item[iii)]{Finally, one determines the coupling $\omega$ in the short-range
    potential from the solution of the matching equation at $O(\delta)$.}
\end{itemize}

The two-channel hadronic potential that is obtained in a result of the
matching condition reproduces, by construction, the relativistic amplitudes
at threshold. The $\pi^+\pi^-\rightarrow\pi^0\pi^0$ amplitude is reproduced 
at $O(\delta)$, whereas $\pi^+\pi^-\rightarrow\pi^+\pi^-$,
$\pi^0\pi^0\rightarrow\pi^0\pi^0$ at $O(1)$ (no isospin breaking). Moreover,
the potential reproduces the decay width of the $\pi^+\pi^-$ ground state at
$O(\delta^{9/2})$.

Note that, if one is willing to reproduce amplitudes in all three channels at
$O(\delta)$ at threshold, as well as to describe simultaneously the energy
shift and the decay of the $\pi^+\pi^-$ bound state, the simple
ansatz~(\ref{lambda_AB}) does not suffice. The possible generalization of 
this ansatz is given by
\eq\label{abc}
u_1=\hat O\, u_0+u_0\,\hat O\, ,\quad\quad
\hat O\doteq\omega_1{\bf 1}+\omega_2\sigma_1+\omega_3\sigma_3\, ,
\en
where ${\bf 1}$ denotes the unit $2\times 2$ matrix, $\sigma_i$ are Pauli
matrices, and real parameters $\omega_1,~\omega_2,~\omega_3$ are the
couplings in the short-range potential at $O(\delta)$. The generalization of
the matching condition at $O(\delta)$ to the elastic channels with the use of
the above ansatz is straightforward. 

\section{Comparison to the existing potential models}
\setcounter{equation}{0}
\label{sec:Comparison}

As was discussed in the Introduction,
the calculations of the observables of hadronic atoms carried out within the
potential approach (see e.g.~\cite{Rasche,Rasche-new,Rasche-talk,Sigg}) often lead to the results
which are in a pronounced disagreement with those obtained on the basis
of ChPT. In addition, these models are sensitive to the particular choice
of the interaction potential, and/or to the choice of the unperturbed Green
function in the Lippmann-Schwinger equation - the different choice may
sometimes lead to the dramatic consequences~\cite{Rasche,Rasche-new,Rasche-talk}. The reason for this
is now clear: the existing calculations use the potential which is not matched
to ChPT in the isospin-breaking phase. Instead, e.g. in Refs.~\cite{Rasche,Rasche-new,Rasche-talk}
the strong potential is matched to ChPT phase shifts 
up to $500~{\rm MeV}$ that, in our
terminology, corresponds to the matching of the parameters of the
effective-range expansion in the strong sector at a high order (see
Section~\ref{sec:strong}). In the contrary, the isospin-breaking part of the
short-range potential is put to zero by hand: it is assumed that the entire
isospin-breaking effect in QCD is due to the Coulomb interactions, kinematical
effects due to the mass difference between charged and neutral particles, etc.
However, it is clear from our construction that such an assumption already
at the next-to-leading order ignores some contributions which are present in
QCD and are contained in the isospin-breaking part of the amplitude 
${\cal A}_1^R$ (Eq.~(\ref{exp-1c})). Below, we list several of these
contributions. 

{\bf 1.} Scattering amplitudes in QCD explicitly depend on quark masses - this
dependence is governed by the chiral symmetry. For demonstrative reason, let
us consider the tree-level scattering amplitude
$\pi^+\pi^-\rightarrow\pi^0\pi^0$ in ChPT (see Ref.~\cite{Bern4} for more
details)
\eq\label{+-00}
T^R_{\pi^+\pi^-\rightarrow\pi^0\pi^0}=-\frac{s-2\hat m B}{F^2}\, ,
\en
where $s$ is the Mandelstam variable, $\hat m=\frac{1}{2}(m_u+m_d)$, and the
constants $F$ and $B$ are related to the pion decay constant and the quark
condensate in the chiral limit. At order $p^2$ in ChPT, 
$2\hat mB=M_{\pi^0}^2$. According to our convention, however, the pion mass in
the isospin-symmetric world is equal to $M_{\pi^+}$, and the splitting of the
above scattering amplitude at threshold into the isospin-conserving and
isospin-breaking parts takes the form
\eq\label{mhat}
T^R_{\pi^+\pi^-\rightarrow\pi^0\pi^0}\biggr|_{thr}=
-\frac{3M_{\pi^+}^2}{F^2}-\frac{M_{\pi^+}^2-M_{\pi^0}^2}{F^2}\, ,
\en
where the isospin-breaking correction (second term of Eq.~(\ref{mhat})) arises
due to the explicit dependence of the scattering amplitude (\ref{+-00}) on 
$\hat m$. Furthermore, the relativistic Lagrangian that leads to the
scattering amplitude (\ref{+-00}) in the tree approximation, should also
contain $\hat m$ 
\eq\label{Lpipi}
{\cal L}_{\pi\pi}^{str}=-\frac{\hat mB}{F^2}\,
\varphi^+\varphi^-\varphi^0\varphi^0+{\rm{terms~with~derivatives}}\, .
\en 
On the language of the potential theory, this would correspond to the
short-range potential with the coupling proportional to $\hat m$. This
coupling changes when one goes from the isospin-symmetric world with 
$2\hat m'B=M_{\pi^+}^2$ to the real world $2\hat mB=M_{\pi^0}^2$
(the constant $B$ is chosen to be the same). Put differently, in order to take
the above effect into account, the short-range potential should contain the
isospin-breaking piece. Since it is not the case in the existing potential
models, this effect is missing there.

{\bf 2.} Direct quark-photon interactions in QCD that occur at a QCD scale
(around $1~{\rm GeV}$), correspond to local vertices in ChPT (see
Fig.~\ref{fig:k_i}). The typical four-pion Lagrangian has the form similar
to~(\ref{Lpipi}) 
\eq\label{Lki}
{\cal L}_{\pi\pi}^{q\gamma}=e^2k_i\,
\varphi^+\varphi^-\varphi^0\varphi^0+{\rm{terms~with~derivatives}}\, ,
\en 
where $k_i$ are the so-called electromagnetic LECs.
The potential which corresponds to this interaction, obviously vanishes in the
isospin limit $e=0$ - consequently, it is not included in the existing
potential models as well.

{\bf 3.} Virtual photon corrections like ones indicated in Ref.~\cite{Fettes}
may lead to the serious discrepancy with the predictions of the potential model.
The counterpart of such a diagram for $\pi\pi$ scattering process is depicted
in Fig.~\ref{fig:Fettes}. It is clear that, at low energy, the contribution of
this diagram is regular in kinematical variables. Consequently, it should be
included in the isospin-breaking part of the short-range potential.
Since the existing potentials are assumed to be isospin-symmetric, they miss
this particular contribution as well.

\vspace*{.2cm}

There is a number of other effects which are not properly included in the
potential models (e.g. the kinematical mass-splitting effect in $t$ and $u$
channels), as well as the effects which are correctly treated
(e.g. resummation of Coulomb ladders). As a general rule, the potential model
gives a reliable prediction of the isospin-breaking effects if and only if
the potential is matched to ChPT in the isospin-breaking phase. We note that
the effects that were discussed in this Section, are not by all means small.
In fact, they form the bulk of the total
isospin-breaking corrections both in the $\pi^+\pi^-$ atom width and the
$\pi^-p$ atom energy shift. It is worth to mention that, in addition to the
isospin-breaking effects, the matching takes care of the differences caused by
choice of the potential and of the free Green function. In particular, there is
no necessity to fit the strong phases up to high energies: it suffices to fit
only the scattering lengths.

From the discussion above, it is clear that, strictly within its range of
applicability, the potential approach does not provide us with a new physical
information as compared to ChPT. In the view of the
extremely complicated analysis of the isospin-breaking corrections in the
scattering processes~\cite{Rasche-scat}, it can be still useful, for the time
being, to construct the potential with full isospin-breaking content of QCD at
threshold. In the absence of systematic analysis carried out on the basis of
ChPT, one may hope that this construction would allow one to
improve the quality of predictions of isospin breaking for the
scattering experiments.

\section{Summary}
\setcounter{equation}{0}
\label{sec:Conclusions}

\begin{itemize}
\item[i)]{On several examples with an increasing level of complexity, we have
    demonstrated the guidelines for the 
    derivation of the short-range hadronic potential from underlying field
    theory of strong and electromagnetic interactions. These potentials, when
    inserted in the conventional Lippmann-Schwinger equation, reproduce, by
    construction, the threshold scattering amplitudes and observables of
    hadronic bound states in the first non-leading order in isospin breaking
    parameter(s).} 
\item[ii)]{According to the viewpoint adopted in the present paper, the
    conventional short-range potentials are considered as a mere
    regularization of the singular pointlike interactions which describe
    low-energy interactions of hadrons in the field theory. For this reason,
    the shape of the potential does not bear a physical
    information. Couplings in the potential are determined from matching of
    the scattering amplitudes in the potential approach and in
    the underlying field theory, both expended in powers of the CM momentum
    squared ${\bf p}^2$. Performing the matching at higher order in
    ${\bf p}^2$, one arrives at a more accurate description of both the
    scattering amplitudes and bound-state observables in the potential
    approach. Apart from the above matching condition, no further restriction
    is imposed on the potentials.} 
\item[iii)]{The derivation of the potential for the description of bound
    states is based on the universality, which states that the properties of
    bound states are the same in the potential scattering approach and field
    theory, once the threshold scattering amplitudes are the same. Due to the
    universality, one may carry out the matching in the scattering sector,
    where the perturbation expansion in $\alpha$ works.}
\item[iv)]{The reason why the results of calculations for the observables of
    hadronic atoms carried out within the potential
    approach~\cite{Rasche,Rasche-new,Rasche-talk,Sigg} generally differ from those obtained in ChPT, is now crystal clear. In order to agree with the latter,
    the potentials should be matched to QCD in the isospin-breaking phase. 
The matching, in general, generates a nonzero 
    isospin-breaking part of the short-range hadronic potential.
Furthermore, it may turn
    out that the prediction for the isospin-breaking corrections to the atom
    observables in the potential approach is close to that of ChPT~\cite{Rasche-talk}. This means
    that the isospin-breaking part of the short-range potential (only in this
    particular hadronic channel) constructed
    through the matching procedure, is very small.}
\item[v)]{It is obvious that, in the context of hadronic atoms, the potential
    which we have 
    constructed does not contain new physical information as compared to
    already available solution of the problem on the basis of effective
    chiral Lagrangians. However, it is interesting to study, whether the
    constraints imposed 
    on the potential from the matching of the amplitudes at threshold, can be
    useful in the analysis of the scattering data along the lines similar to
    those from Refs.~\cite{Rasche-scat}.} 
\end{itemize}

\section*{Acknowledgments}
We are grateful to J. Gasser for the current interest in the work and useful
suggestions. We thank 
A. Badertscher, G. Dvali, 
H.-J. Leisi, H. Leutwyler, B. Loiseau, L.L. Nemenov, 
M. Pavan, G.C. Oades, G. Rasche, M.E. Sainio, and J. Schacher
for interesting discussions.
The work was fulfilled while E.L. visited the University of Bern. 
This work was supported in part by the Swiss National Science
Foundation, and by TMR, BBW-Contract No. 97.0131  and  EC-Contract
No. ERBFMRX-CT980169 (EURODA$\Phi$NE), and by SCOPES Project No. 7UZPJ65677.

\appendix

\renewcommand{\thesection}{\Alph{section}}
\renewcommand{\theequation}{\Alph{section}\arabic{equation}}
\setcounter{section}{0}
\section{Behavior of the scattering matrix at a small momenta}
\setcounter{equation}{0}
\label{app:phi}

In this Appendix, we consider the behavior of the scattering matrix at a 
small momenta in the potential scattering theory with a short-range potential
$u$. The $S$-wave potential in momentum space is given via the Fourier
transform 
\eq\label{A1}
u(q,p)=16\pi^2\int\limits_{0}^{\infty}r_1\,d\,r_1
\int\limits_{0}^{\infty}r_2\,d\,r_2\,\frac{\sin(qr_1)}{q}\,
\frac{\sin(pr_2)}{p}\,u(r_1,r_2)\, ,
\en
where $u(r_1,r_2)$ is assumed to be short-ranged. The expansion of the
right-hand side of Eq.~(\ref{A1}) in powers of $q$ contains only even powers
of $q$, provided the integrals in $r_1$ which emerge after the expansion of
the integrand, are convergent. In particular,
\eq
\lim\limits_{q^2\to 0}\, \frac{u(q,p)-u(0,p)}{q^2}=-\frac{8\pi^2}{3}\int
\limits_0^{\infty}r_1^4\,dr_1\int\limits_0^{\infty}r_2\,dr_2\,
\frac{\sin (pr_2)}{p}\,u(r_1,r_2)\, ,
\en
if the integral in the right-hand side is convergent. We assume that our
short-range potential satisfies this requirement.

The scattering matrix satisfies the Lippmann-Schwinger equation
\eq
t(E;q,p)=u(q,p)+\frac{1}{2\pi^2}\int\limits_0^{\infty}\frac{k^2\,dk}{E-k^2/M}\,
u(q,k)\,t(E;k,p)\, .
\en

Using this equation, it is immediately seen that the scattering matrix has 
the similar behavior at small $q$, as the potential
\eq
\lim\limits_{q\to 0}\,\frac{t(E;q,p)-t(E;0,p)}{q^2}<\,\infty\, .
\en
The equation~(\ref{phi0}) directly follows from this equation.

\section{Decay width in the two-channel scattering case}
\setcounter{equation}{0}
\label{app:decay}

The perturbative solution of Eq.~(\ref{potential-pole}) gives
\eq
z-E_0&-&(\Psi_0|\,t^{++}(E_0)\,|\Psi_0) -
(\Psi_0|\,t^{++}(E_0)\,{\bf g}^{1-C}(E_0)\,t^{++}(E_0)\,|\Psi_0)
\nonumber\\[3mm]
&-&(\Psi_0|\,t^{++}(E_0)\,{\bf g}^{n-C}(E_0)\,t^{++}(E_0)\,|\Psi_0) +
\cdots=0\, .
\en
The separate contributions in this equation correspond to splitting of
the Coulomb Green function into the no-photon, one-photon, and
many-photon pieces. 

In the evaluation of the matrix element with no photons, we are again
confronted with the necessity of the analytic continuation of the scattering
matrix from the bound-state energy $E_0$ to the scattering threshold $z_+$.
Due to the presence of the unitary cusp in the real part of the scattering
matrix, in a result of such a continuation we get the difference which is
non-vanishing at $O(\delta)$ 
\eq
{\rm Re}\, t^{+0}(E_0;0,k_0(E_0))-{\rm Re}\, {\bar t}^{+0} 
= \frac{\alpha M_{\pi^+}^2}{8\pi} ({\bar t}^{++})^{\star}\, {\bar t}^{+0}
+o(\delta)\, .
\en

The imaginary part of the matrix element with no Coulomb photon exchange is
given by
\eq
{\rm Im}\,(\Psi_0|\,t^{++}(E_0)\,|\Psi_0)&=&
-\frac{M_{\pi^0} k_0(E_0)}{4 \pi}\,
\phi_0^2\left \{ |{\bar t}^{+0}|^2
\right.\nonumber\\[2mm]
&+&\left. \frac{\alpha M_{\pi^+}^2}{4\pi}\,
|{\bar t}^{+0}|^2\,{\rm Re}\,{\bar t}^{++} +
\frac{4 M_{\pi^+}\alpha}{\pi}\,{\rm Re}\,{\bar t}^{+0}\, 
{\cal Q}[t^{+0}]\right\}+ o(\delta^{9/2})\, ,
\en
where
\eq
{\tilde {\cal Q}}[t]=\int\limits_0^{\infty}\frac{dp}{p^2}\,\left (
{\rm Re}\,t^{+0}(z_+;p,k_0(z_+))-{\rm Re}\, {\bar t}^{+0}\right )\, .
\en

The imaginary part of the matrix element with one Coulomb photon exchange is
given by
\eq
&&{\rm Im}(\Psi_0|\,t^{++}(E_0)\,{\bf g}^{1-C}(E_0)\,t^{++}(E_0)\,|\Psi_0)=
\nonumber\\[2mm]
-\frac{M_{\pi^0}k_0(E_0)}{4\pi}\,\phi_0^2 && \left\{
-\frac{\alpha M_{\pi^+}^2}{2\pi}\, |{\bar t}^{+0}|^2
\,{\rm Re}\,{\bar t}^{++}\, \ln\frac{b}{2\gamma}+
\frac{\alpha M_{\pi^+}^3}{2\pi^3}\, {\rm Re}\,{\bar t}^{+0}\,
{\tilde {\cal R}}[b;t]\right\}+o(\delta^{9/2})\, ,
\en
where
\eq
{\tilde {\cal R}}[b;t]&=&\int\limits^{\infty}_{0}\frac{dp
  dk}{pk}\left\{\ln\frac{(p-k)^2}{(p+k)^2}\,
-\ln\frac{(p-k)^2+b^2}{(p+k)^2+b^2}\right\}\nonumber\\[3mm]
&\times&\left [
{\rm Re}\, t^{++}(z_+;0,p)\, {\rm Re}\, t^{+0}(z_+;k,k_0(z_+))-
{\rm Re}\, {\bar t}^{++}\, {\rm Re}\, {\bar t}^{+0} 
\right]\\[3mm]
&+&\int\limits^{\infty}_{0}\frac{dp
dk}{pk}\,\ln\frac{(p-k)^2+b^2}{(p+k)^2+b^2}\,{\rm Re}\, t^{++}(z_+;0,p)\,
{\rm Re}\, t^{+0}(z_+;k,k_0(z_+))\, .\nonumber
\en

Finally, the matrix element containing many-Coulomb photon exchanges, is 
given by the following expression
\eq
{\rm Im}\,(\Psi_0|\,\{t(E_0)\,{\bf g}^{n-C}(E_0)\,t(E_0)\}^{++}\,|\Psi_0)=-
\frac{M_{\pi^0} k_0(E_0)}{4\pi}\,\phi_0^2\left\{-\frac{3\alpha M_{\pi^+}^2}{4\pi}\,
|{\bar t}^{+0}|^2\, {\rm Re}\, {\bar t}^{\,\,++}\right\}
\en

Adding things together, and evaluating the decay width 
$\Gamma=-2{\rm Im}\, z$, we arrive at Eq.~(\ref{width}) 
where the quantity ${\cal A}$ is given by Eq.~(\ref{A-decay}).

\section{Matching of the potential in the two-channel scattering case}
\setcounter{equation}{0}
\label{app:potential}

In order to obtain the explicit form of the functional $\tilde{\cal F}[t_0]$
in the matching condition for the potential~(\ref{1delta}), 
we first evaluate the integrals containing 
${\tilde {\bf g}}_0-{\bar {\bf g}}_0$, which appear in the r.h.s. of
Eq.~(\ref{split})  
\eq
&&{\rm Re}\left\{ t_0({\tilde {\bf g}}_0-{\bar {\bf g}}_0) t_0\right\}^{+0}
\biggr |_{\rm thr}=\frac{M_{\pi^+}}{4\pi^2}\,\kappa\left \{{\cal G}_1[t_0]-
2 M_{\pi^+}^2 {\cal Q}_1[t_0]\right \}+o(\delta)\, ,
\nonumber\\[2mm]
&&{\rm Re}\left\{ t_0({\tilde {\bf g}}_0-{\bar {\bf g}}_0) t_0
({\tilde {\bf g}}_0-{\bar {\bf g}}_0) t_0\right\}_{+0}
\biggr |_{\rm thr}=-\frac{M_{\pi^+}^4}{16\pi^2}\,\kappa\, 
{\bar t}_0^{\,\,+0}\, ({\bar t}_0^{\,\,00})^2+o(\delta)\, ,
\en
where
${\bar t}_0^{\,\,AB}\doteq t_0^{AB}(z_+;0,0)$ the quantity $\kappa$ is defined
after formula (\ref{K-pipi}), and
\eq
{\cal G}_1[t_0]&=&\int\limits_0^{\infty}dp\, t_0^{+0}(z_+;0,p)
\, t_0^{00}(z_+;p,0)\, ,\nonumber\\[2mm]
{\cal Q}_1[t_0]&=&\int\limits_0^{\infty}\frac{dp}{p^2}\,
\left [t_0^{+0}(z_+;0,p)\, t_0^{00}(z_+;p,0)- {\bar t}_0^{\,\,+0}\,
{\bar t}_0^{\,\,00}\right ]\, .
\en

Substituting these expressions into the matching condition for the amplitude
in the charge-exchange channel, we finally obtain
\eq\label{1deltaC}
\tilde{\cal F}[t_0]
&=&\frac{M_{\pi^+}}{4\pi^2}\,\kappa\,\left\{{\cal G}_1[t_0]-
2M_{\pi^+}^2\,{\cal Q}_1[t_0]\right\}
-\frac{M_{\pi^+}^4}{16\pi^2}\,\kappa\,{\bar t}_0^{+0}({\bar t}_0^{00})^2
+\frac{2\alpha M_{\pi^+}}{\pi}\, \tilde {\cal Q}[t_0]\nonumber\\[2mm]
&-&\frac{\alpha M_{\pi^+}^2}{4\pi}\,\ln\frac{b}{M_{\pi^+}}\,
{\bar t}_0^{+0}\,{\bar t}_0^{\, ++}+
\frac{\alpha M_{\pi^+}^2}{4\pi^3}\,{\tilde {\cal R}}[b;t_0]+o(\delta)\, .
\en
where the functionals $\tilde{\cal Q}[t_0]$ and $\tilde{\cal R}[b;t_0]$
are defined in Appendix~\ref{app:decay}.

\vspace*{2.5cm}

\begin{center}
{\bf FIGURE CAPTIONS}
\end{center}

\begin{figure}
\vspace*{7cm}
\hspace*{1.5cm}
\includegraphics{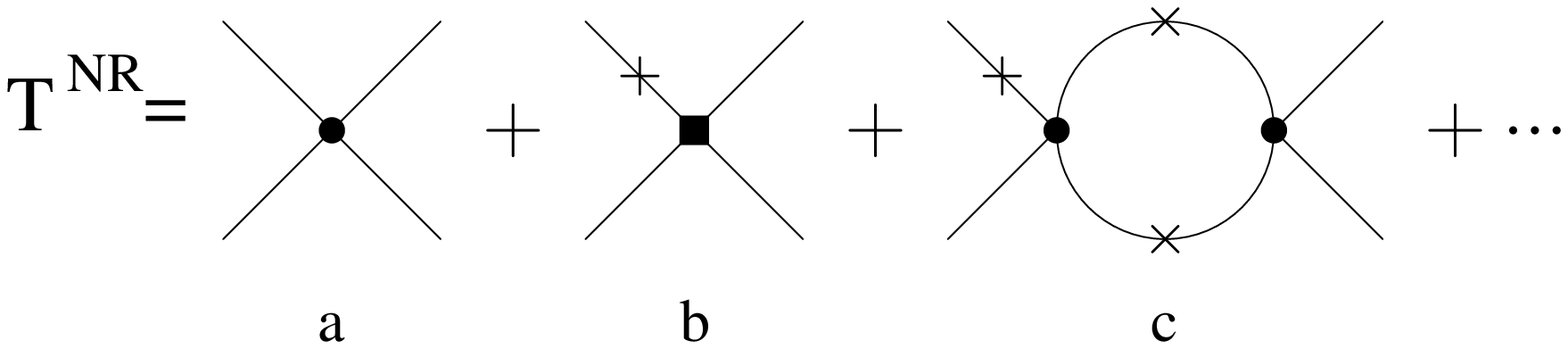}
\vspace*{-2.5cm}
\caption{}\label{fig:strongbubbles}
\end{figure}

\noindent {\bf FIG.~\ref{fig:strongbubbles}.}
Representative diagrams that contribute to the non-relativistic scattering
amplitude in the purely strong case: a) tree diagram with the
non-derivative coupling $g_0$; b) tree diagram with the derivative coupling
in the four-pion vertex ($g_1$ or $g_2$), and with mass insertions;
c) one strong bubble with non-derivative couplings and with mass insertions
in the external and internal lines. The filled circle and filled square 
denote the non-derivative and derivative vertices, respectively, and crosses
stand for the mass insertions.
\vspace*{.3cm}

\begin{figure}
\vspace*{15cm}
\hspace*{1.5cm}
\includegraphics{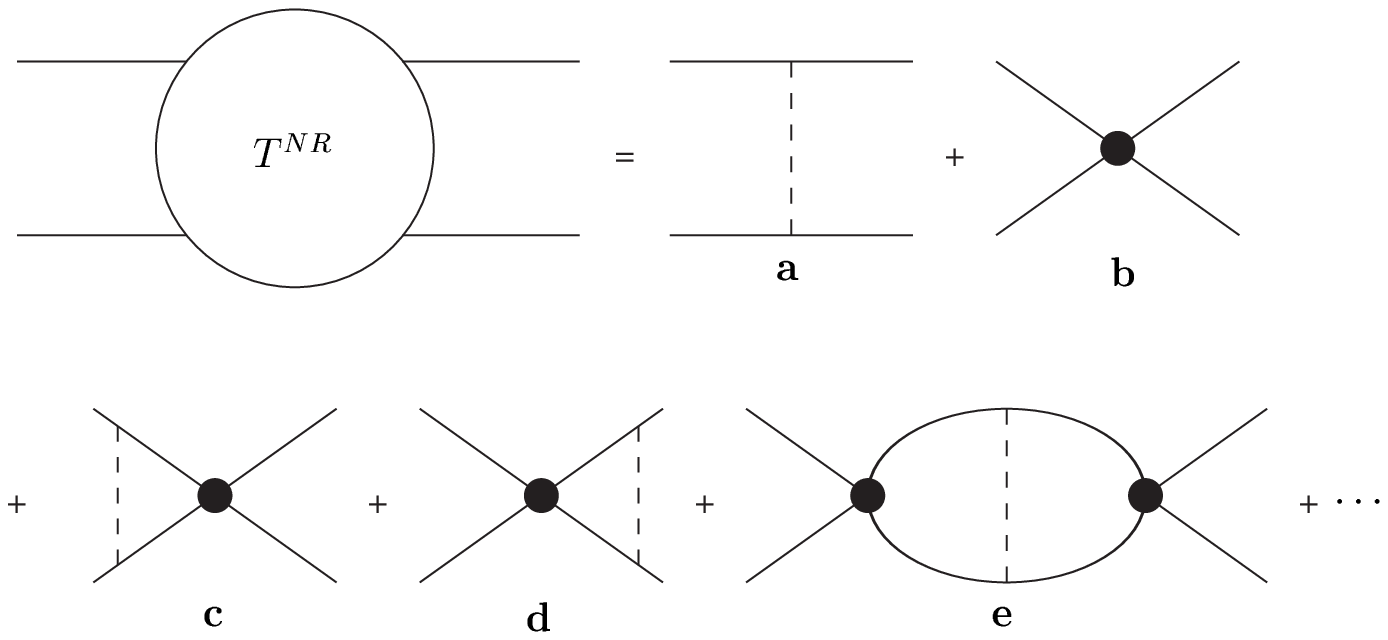}
\vspace*{-8cm}
\caption{}\label{fig:NR-1c}
\end{figure}

\noindent {\bf FIG.~\ref{fig:NR-1c}.}
Building blocks for the non-relativistic scattering amplitude at $O(\alpha)$:
a) one Coulomb photon exchange; b) strong 4-particle vertex; c),d) exchange
of Coulomb photon in the initial and final states; e) two-loop diagram with the
exchange of Coulomb photon in the intermediate state.

\vspace*{.3cm}

\begin{figure}
\vspace*{15cm}
\hspace*{-.20cm}
\includegraphics{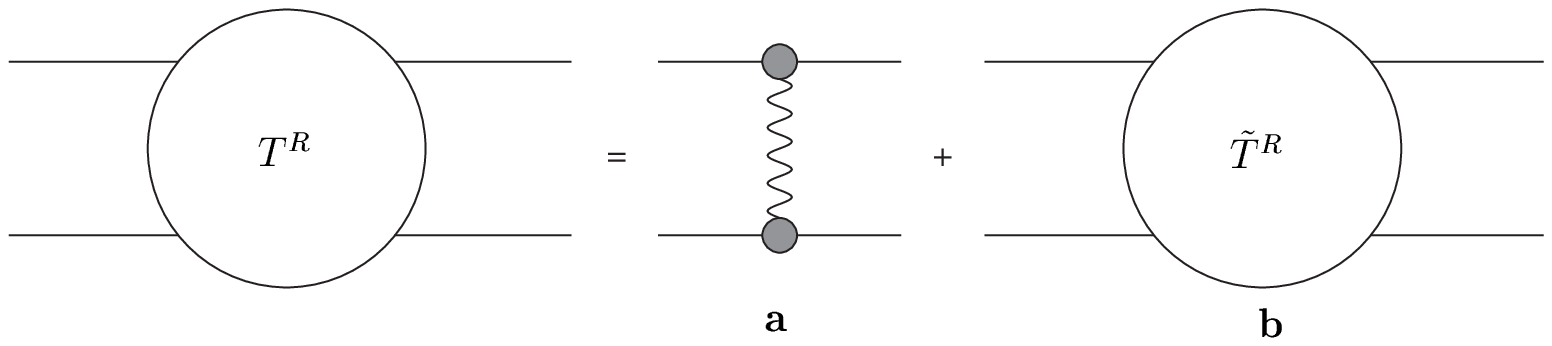}
\vspace*{-10cm}
\caption{}\label{fig:R-1c}
\end{figure}

\noindent {\bf FIG.~\ref{fig:R-1c}.}
Decomposition of the relativistic scattering amplitude into one-photon
exchange part (a) and the truncated amplitude (b).

\vspace*{.3cm}

\begin{figure}
\vspace*{14cm}
\hspace*{.5cm}
\includegraphics{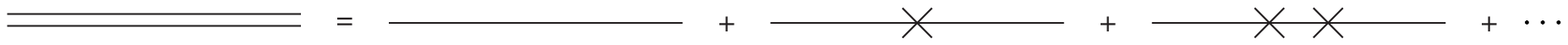}
\vspace*{-9.5cm}
\caption{}\label{fig:sum}
\end{figure}

\noindent {\bf FIG.~\ref{fig:sum}.}
Summation of the mass insertions in the non-relativistic pion propagator.

\vspace*{.3cm}

\begin{figure}
\vspace*{15cm}
\hspace*{-.1cm}
\includegraphics{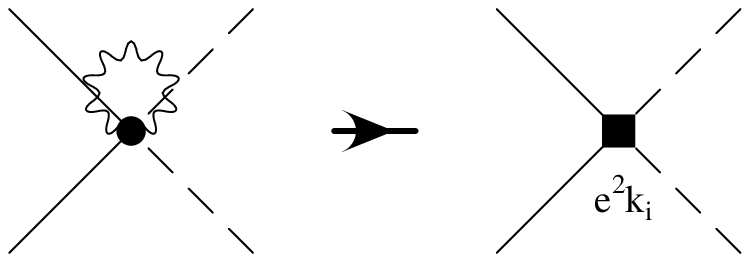}
\vspace*{-9cm}
\caption{}\label{fig:k_i}
\end{figure}

\noindent {\bf FIG.~\ref{fig:k_i}.}
Direct quark-photon interactions in QCD and in ChPT.
Solid and dashed lines denote $\pi^\pm$ and $\pi^0$, respectively.

\vspace*{.3cm}

\begin{figure}
\vspace*{15cm}
\hspace*{2.7cm}
\includegraphics{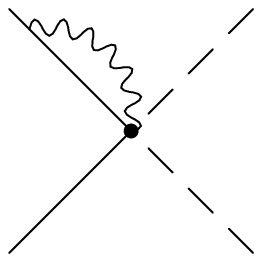}
\vspace*{-9cm}
\caption{}\label{fig:Fettes}
\end{figure}

\noindent {\bf FIG.~\ref{fig:Fettes}.}
A particular diagram corresonding to the  virtual photon correction to the
scattering amplitude $\pi^+\pi^-\rightarrow\pi^0\pi^0$. 
Solid and dashed lines denote $\pi^\pm$ and $\pi^0$, respectively.
Filled dot stands for the non-minimal photon coupling to four pions.

\newpage

\end{document}